\pdfoutput=1
\documentclass[a4paper,fleqn,usenatbib]{mnras}
\usepackage[T1]{fontenc}
\usepackage{ae,aecompl}
\usepackage{graphicx}	% Including figure files
\usepackage{amsmath}	% Advanced maths commands
\usepackage{amssymb}	% Extra maths symbols

\title[Predicting extragalactic distance errors]{Predicting extragalactic distance errors using Bayesian inference in multi-measurement catalogs}

\author[G. Chaparro-Molano et al.]{Germ\'an Chaparro-Molano$^{1}$\thanks{E-mail: gchaparrom@ecci.edu.co},
Juan Carlos Cuervo,
Oscar Alberto Restrepo Gait\'an$^{1,2}$, \newauthor
Sergio Torres Arzay\'{u}s$^{3}$
\\
$^{1}$Vicerrector\'ia de Investigaci\'on, Universidad ECCI, Bogot\'a, Colombia\\
$^{2}$Radio Astronomy Instrumentation Group, Universidad de Chile, Santiago de Chile, Chile\\
$^{3}$Centro Internacional de F\'isica, Bogot\'a, Colombia
}

\date{Accepted XXX. Received YYY; in original form ZZZ}

\pubyear{2018}

\begin{document}
\label{firstpage}
\pagerange{\pageref{firstpage}--\pageref{lastpage}}
\maketitle

\begin{abstract}
We propose the use of robust, Bayesian methods for estimating extragalactic distance errors in multi-measurement catalogs. We seek to improve upon the more commonly used frequentist propagation-of-error methods, as they fail to explain both the scatter between different measurements and the effects of skewness in the metric distance probability distribution. For individual galaxies, the most transparent way to assess the variance of redshift independent distances is to directly sample the posterior probability distribution obtained from the mixture of reported measurements. However, sampling the posterior can be cumbersome for catalog-wide precision cosmology applications. We compare the performance of frequentist methods versus our proposed measures for estimating the true variance of the metric distance probability distribution. We provide pre-computed distance error data tables for galaxies in 3 catalogs: NED-D, HyperLEDA, and Cosmicflows-3. Additionally, we develop a Bayesian model that considers systematic and random effects in the estimation of errors for Tully-Fisher relation (TF) derived distances in NED-D. We validate this model with a Bayesian $p$-value computed using the Freeman-Tukey discrepancy measure as a posterior predictive check. We are then able to predict distance errors for 884 galaxies in the NED-D catalog and 203 galaxies in the HyperLEDA catalog which do not report TF distance modulus errors. Our goal is that our estimated and predicted errors are used in catalog-wide applications that require acknowledging the true variance of extragalactic distance measurements.
\end{abstract}


%%%%%%%%%%%%%%%%%%%%%%%%%%%%%%%%%%%%%%%%%%%%%%%%%%

%%%%%%%%%%%%%%%%% BODY OF PAPER %%%%%%%%%%%%%%%%%%

\section{Introduction}

Understanding the uncertainties in redshift-independent extragalactic distance measurements is absolutely necessary before reporting statistically sound conclusions regarding the structure of the local universe \citep{void,locunivcf,nongauss,6df,localunipv,said,gg3500}, large scale structure \citep{anishub,gallargescale,morphanis,tecciencia,bayesh}, and events like transient gravitational wave detections \citep{gwgallist}. Hubble constant estimations have been using increasingly sophisticated statistical tools for primary distance determination methods, such as SNIa \citep{ridsn,unity,hubsn2018}, Cepheids \citet{hubngc} or both \citep{riess}. Although most estimates of the Hubble constant use Cepheid calibration for calibrating secondary methods \citep{hubunc,huborig,hub2010}, \citet{noceph} have explored changes in Hubble constant estimation using the Tully-Fisher relation (TF) without Cepheid calibration. Secondary methods for extragalactic distance determination like the TF relation, or the Fundamental Plane (FP) have recently become more precise thanks to increasing volumes of data from surveys like 6dF \citep{6df} and 2MASS \citep{2mass,tf07dist} together with Spitzer data \citep{sorce}, along with improved statistical methods \citep{precisetf}. \\

As of 2018, three multi-measurement catalogs including a substantial amount of redshift-independent extragalactic distance measurements have been released: HyperLEDA \citep{hyperleda}, NED-D \citep{ned07,ned}, and Cosmicflows-3 \citep{cosmicflows}. HyperLEDA includes a homogenized catalog for extragalactic distances in the nearby universe, with 12866 distance measurements for 518 galaxies to date. NED-D is the NASA/IPAC Extragalactic Distance catalog of Redshift-Independent Distances, which compiles 326850 distance measurements for 183062 galaxies in its 2018 version. Here, $\sim1800$ galaxies ($\sim1$\%) have more than 13 distance measurements, and $~180$ galaxies ($\sim0.1$\%) have distance measurements using more than 6 different methods. Cosmicflows-3 is the most up-to-date catalog, which reports distance measurements for 10616 galaxies (all of which include errors) using up to four distance determination methods, calibrated with supernova luminosities. However, unlike HyperLEDA or NED-D, Cosmicflows-3 only reports the latest distance measurement for each method. In HyperLEDA, NED-D and Cosmicflows-3 errors are reported as one standard deviation from the reported distance modulus. Treatment of errors for combining distance moduli across methods or across measurements is suggested by \citet{ned07} and \citet{cosmicflows} to be based on weighted estimates such as the uncertainty of the weighted mean, albeit with caution partly due to the heterogeneous origin of the compiled data and partly due to Malmquist bias. In the case of NED-D, this is additionally complicated by the fact that many errors are not reported or are reported as zero. In fact, the TF relation method has the largest number of galaxies with non-reported distance modulus errors (884 to date). Even though extragalactic distances measured using the TF relation were originally reported to have a relative error in distance modulus of $10-20$\% \citep{tforig}, we consider that this conservative estimate can be improved upon by using a predictive model based on the distance error of galaxies that use the same distance determination method. This requires a robust estimation of the variance of extragalactic distances based on the available data.\\

For many galaxies in all three catalogs, the random error for each distance modulus measurement $\epsilon_i$ (for $i=1,...,N$, for $N$ distance measurements per galaxy) is not representative of the scatter across measurements, even when considering the same method for determining distances. In addition, distance modulus distributions for each measurement (which are assumed to be Gaussian) are transformed to log-normal distributions in metric distance space. This can introduce a significant bias in peculiar velocity studies for large-scale structure studies \citep{lognormal}. We improve upon previous methods by robustly estimating the underlying variance across measurements and distance determination methods. To do this, we measure the 84th and 16th percentiles, and the median absolute deviation of the bootstrap-sampled posterior probability distribution of each extragalactic distance \citep{chaparro18}. We compare our results to other more commonly used frequentist methods, such as the weighted estimates mentioned above, and we produce pre-computed data tables for the three catalogs mentioned above, which can be found in the repository for this work at \texttt{https://github.com/saint-germain/errorprediction}. We then perform a Bayesian analysis of the systematics and randomness of the computed errors in the NED-D catalog for TF relation derived distances. From this analysis we build predictive models for the estimation of errors and evaluate them by performing posterior predictive checks using a discrepancy measure-derived Bayesian ``$p$-value'' \citep{gelmanppd}. Furthermore, we make predictions for the 884 galaxies in the NED-D catalog and the 203 galaxies in the HyperLEDA catalog whose distances were measured using the TF relation but have non-reported errors. Inference based on Bayesian posterior predictive checks has been advocated for in \citet{gelman2003} and \citet{ppcinf}.\\

We organize this paper as follows. In Section~\ref{sec:post} we talk about the posterior distribution of distance for individual galaxies and set up methods for measuring its variance. In Section~\ref{sec:comp} we make a comparison between the proposed variance estimation methods, and in Section~\ref{sec:predbay} we propose and evaluate predictive Bayesian models for two robust methods of error estimation, and we summarize our work in the Conclusions section. The appendix includes a description and brief analysis of extragalactic distance error data tables pre-computed with the methods described in this paper for the HyperLEDA, Cosmicflows-3 and NED-D catalogs.

\section{Estimation of extragalactic distance errors}
\label{sec:post} 
 
The best approach to consider the effects of random and systematic errors in catalog-wide, multi-method distance analyses is to directly sample the posterior probability distribution of each extragalactic distance. This can be achieved by drawing distance modulus samples from $P(\mu)$, which is the unweighted mixture of normal distributions corresponding to each distance modulus measurement $\mu_i$,
\[\mu\sim\sum_i^N \mathcal{N}(\mu_i,\epsilon_i^2)\ ,\]
and then converting to metric distance,
\[D=10^{\frac{\mu}{5}+1}\ .\]
Therefore,
\begin{equation}\label{eqn:mix}
D_G\sim\sum_i^N\mathrm{lognormal}(M_i,\sigma_{M_i}^2)\ .
\end{equation}
Here $M_i=\ln D_i$ and $\sigma_{M_i}=\epsilon_i\cdot\ln10$.

\subsection{Estimating the variance of $P(D_G)$}
\label{sec:meth} 

Although directly sampling the distribution of $D_G$ is the most transparent way to acknowledge the true variance of distance measurements, it is not a very efficient way to achieve a standardized treatment of errors. One simple measure of the variance of $D_G$ that acknowledges the possible skewness of the distribution is to take the median 16th and 84th percentile of 10k bootstrap samples of the distribution of $D_G$, e.g. one bootstrap sample corresponds to $N$ draws, one from each reported measurement. In our pre-computed error tables we report these quantities as \texttt{Dmin} and \texttt{Dmax}, respectively.\\

It can be even more convenient to treat each extragalactic metric distance $D_G$ as a normal random variable with a single-valued $\sigma_D$ as a measure of the uncertainty in the estimation of an extragalactic distance,
\begin{equation}\label{eqn:norm}
D_G\sim \mathcal{N}(D,\sigma_D^2)
\end{equation}
For this reason we compare four methods for estimating the $D,\,\sigma_D$ pair. Two of these methods (H, M) use robust measures of the distribution of each extragalactic distance, and the other two (P, Q) use measures based on propagation of errors.\\

Methods H and M, which are the methods we propose to robustly estimate $\sigma_D$ in equation \ref{eqn:norm} are based on measuring the median and variance of repeated bootstrap samples from the distribution of $D_G$ (equation \ref{eqn:mix}) as mentioned in the previous section. Method H takes $D$ as the median of the bootstrap samples and $\sigma_D$ as the half-distance (H) between  the 84th and 16th percentiles of 10k bootstrap samples. We consider this to be the method which most faithfully measures the variance regardless of the shape of the posterior distribution. Method M takes $D$ as the median of the bootstrap samples and $\sigma_D$ as the median absolute deviation (MAD) of the bootstrap samples. This method is better suited for avoiding the effects of outliers.\\

The other two methods (P,Q) considered here are based on commonly used frequentist estimates of the distance error. Method P consists on calculating $D$ from the weighted mean distance modulus $\bar{\mu}^*$ with weights $w_i=\epsilon_i^{-2}$. $\sigma_D$ is calculated by propagation (P) of measurement errors  i.e. from the uncertainty of the weighted mean \citep{cosmicflows},
\begin{equation}
\sigma_D^P=0.461\,\bar{D}^*\,\left(\sum_i^Nw_i\right)^{-1/2} \ ,
\end{equation}
Method P does not take into account the scatter in distance measurements for single galaxies, which is why it can be convenient to calculate $\sigma_D$ as the sum in quadrature (Q) of the propagated uncertainty of the weighted mean and the propagated unbiased weighted sample variance $\sigma_D^*$:
\begin{equation}
\sigma_D^Q=\left[ \left(\sigma_D^P\right)^2+\Big(\sigma_D^*\Big)^2\right]^{1/2} \ .
\end{equation}
Here $\sigma^*_D$ is calculated as  \citep{wstdev},
\begin{equation}
\sigma^*_D=0.461\,\bar{D}^*\,\sqrt{\frac{N}{N-1.5}\frac{\sum_i^Nw_i(\mu_i-\bar{\mu}^*)^2}{\sum_i^Nw_i}\vphantom{\Biggl(}}\ .
\end{equation}
If the non-robust P and Q methods were truly representative of the variance of the distribution of $D_G$, they should yield similar results as the H or M methods. The following section shows that this is not the case.

\section{Comparison of distance error estimation methods}
\label{sec:comp}

In this section we focus on NED-D distance measurements since it is the largest of the three catalogs considered here. A full discussion of our error estimation method applied to multi-method measurements in the HyperLEDA, NED-D and Cosmicflows-3 is given in the appendix. A repository for this work, including the pre-computed error tables for the HyperLEDA, NED-D and Cosmicflows-3 is located at \texttt{https://github.com/saint-germain/errorprediction}. From here on, when we mention distance measurements in the NED-D catalog, we will be excluding from our analysis measurements that require the target redshift to calculate the distance, as indicated in the \texttt{redshift (z)} field.\\

For galaxies with a number of distance measurements between 2 and 5 (Fig.~\ref{fig:NED}, left) , errors estimated with the the quadrature (Q), and median absolute deviation (M) methods show a linear trend with similar slopes that over-predict the variance with respect to the half 84th-16th percentile distance (H) method, whereas the propagation (P) method tends to under-predict the errors. Furthermore, errors estimated using the Q method show a larger dispersion around the linear trend than the H and M methods. Fig.~\ref{fig:NED} (right) shows that the P and Q methods underpredict errors for galaxies with more than 5 distance measurements. 

\subsection{Distance errors in Tully-Fisher relation derived measurements}

Even though our analysis for error estimation can be used to combine distance measurements using different methods for single galaxies, we think that due to method-intrinsic systematics it is more appropriate to separate the analysis by method. Without loss of generality, we now focus on galaxies whose distances have been measured using the Tully-Fisher method in the NED-D catalog because it is the method with the largest number of galaxies without reported measurement errors (884) in the database.  \\

\begin{figure*}

	\includegraphics[scale=0.69]{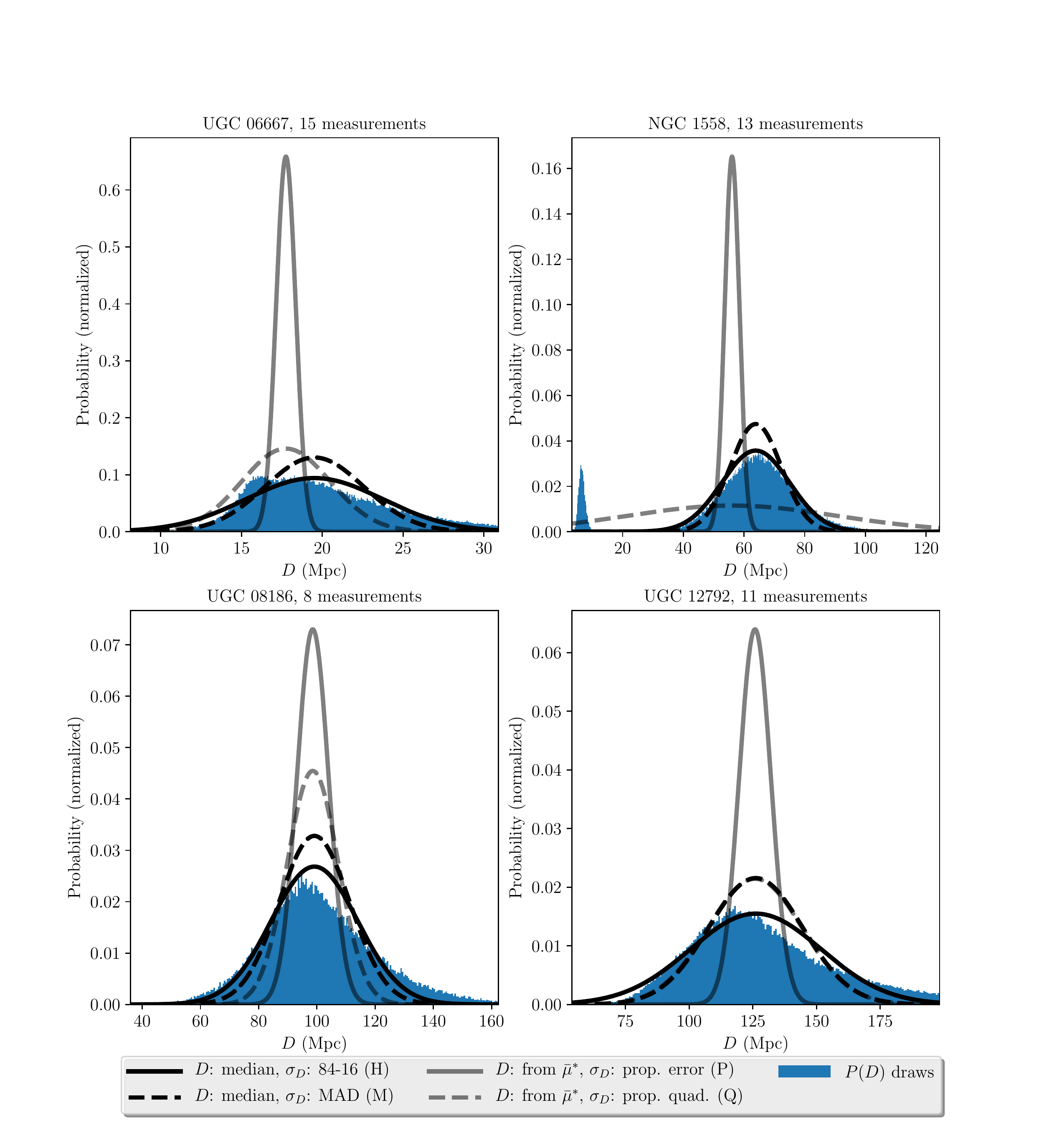}
    \caption{Comparison of four examples of extragalactic distance posterior distribution draws (10000 per measurement) and modeled distributions for UGC 06667, NGC 1558, UGC 08186, and UGC 12792 using the Tully-Fisher Method for distance determination in NED-D. The methods used for approximating the posterior distribution (H, M, P, and Q) are described in the text. }
    \label{fig:comp}
\end{figure*}

Fig.~\ref{fig:comp} shows that for a small but representative sample of galaxies with more than 7 distance measurements, the center and variance of the posterior distribution of each extragalactic distance is best explained using the H method, whereas the less robust P and Q methods under-predict the variance. On the other hand, the M method also under-predicts the variance because it is a robust measure, and thus not as sensitive to outliers as methods P and Q, as seen in the case of NGC 1558 in Fig.~\ref{fig:comp}. For the more symmetrical posterior distribution of UGC 12792, the M and Q methods predict the same center and variance.\\

\begin{figure*}

	\includegraphics[scale=0.69]{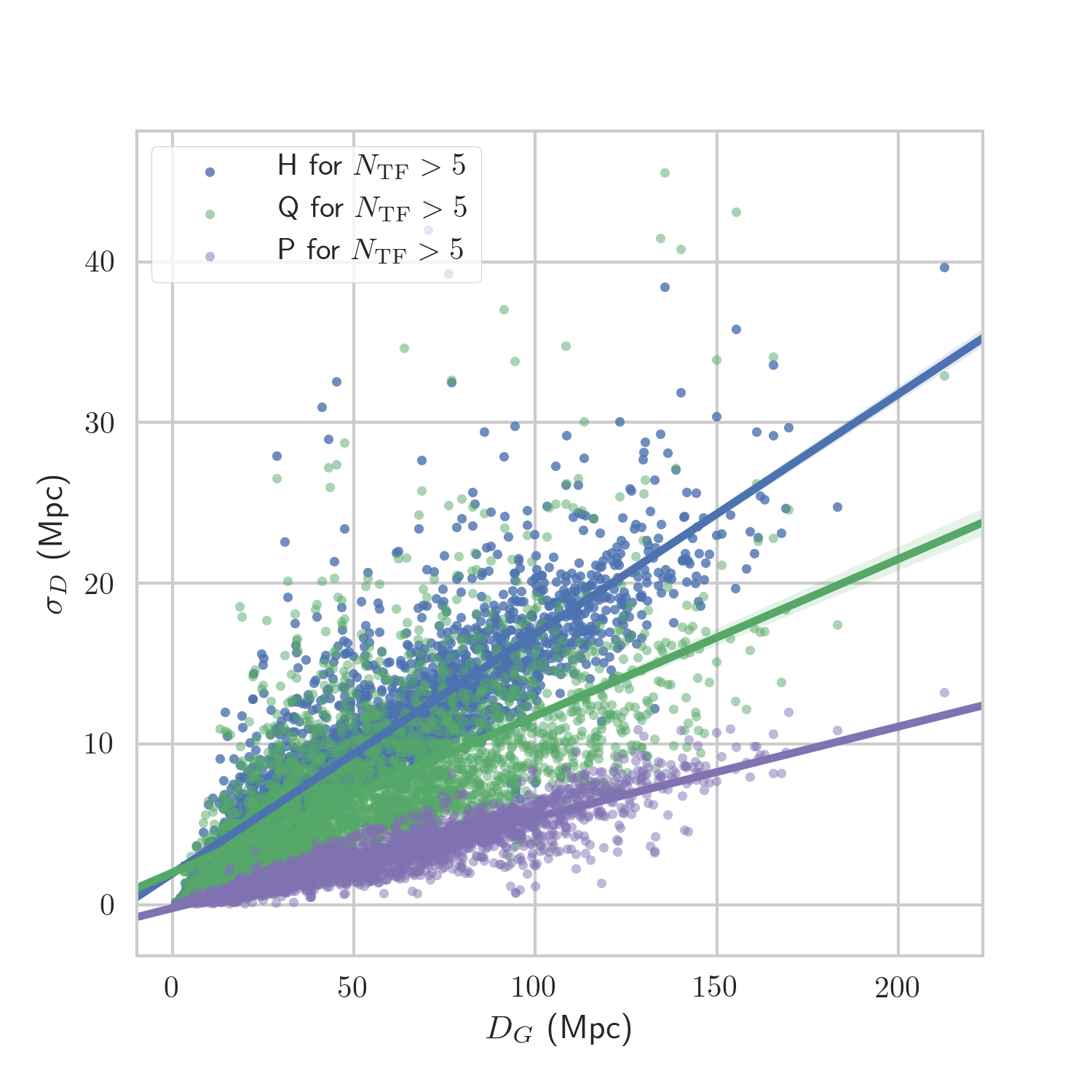}
	\includegraphics[scale=0.69]{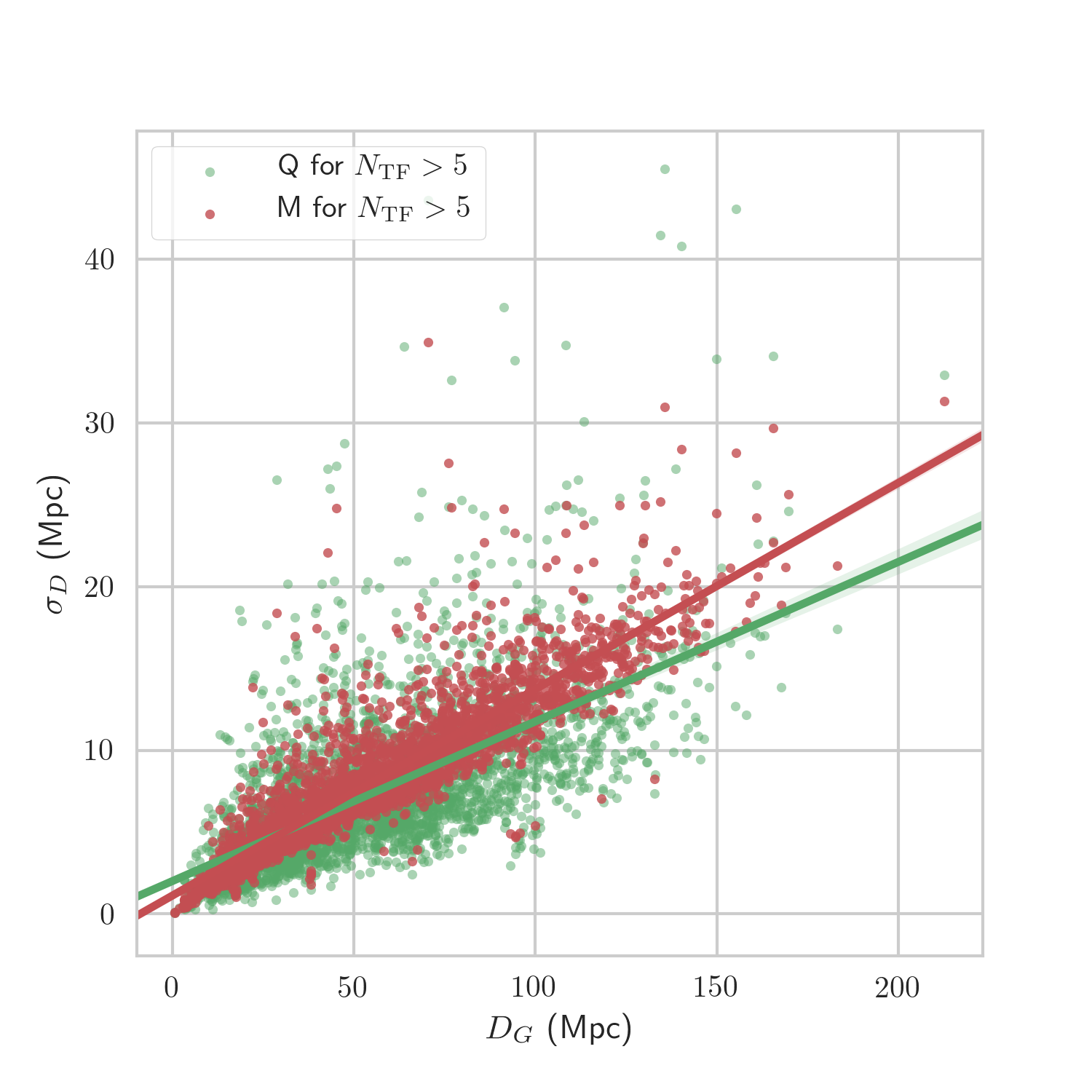}
    \caption{Estimated extragalactic distance errors vs. median extragalactic distance  for galaxies with more than 5 TF distance measurements in NED-D according to the H, Q, P (left) and Q, M (right) error models, showing linear regressions and confidence intervals computed using the \texttt{seaborn.regplot} Python function.}
    \label{fig:hqp-qm}
\end{figure*}

\begin{figure*}
	\includegraphics[scale=0.69]{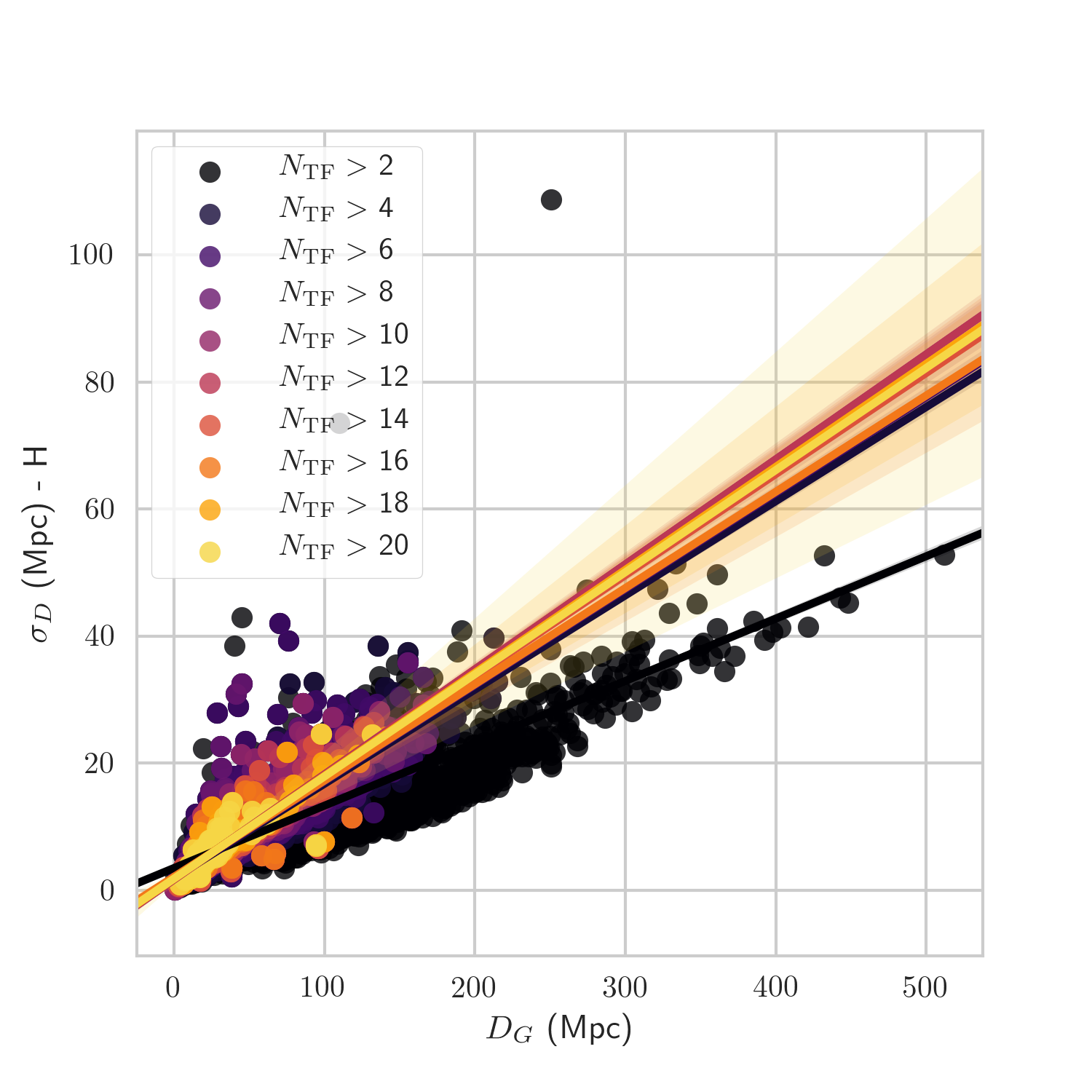}
	\includegraphics[scale=0.69]{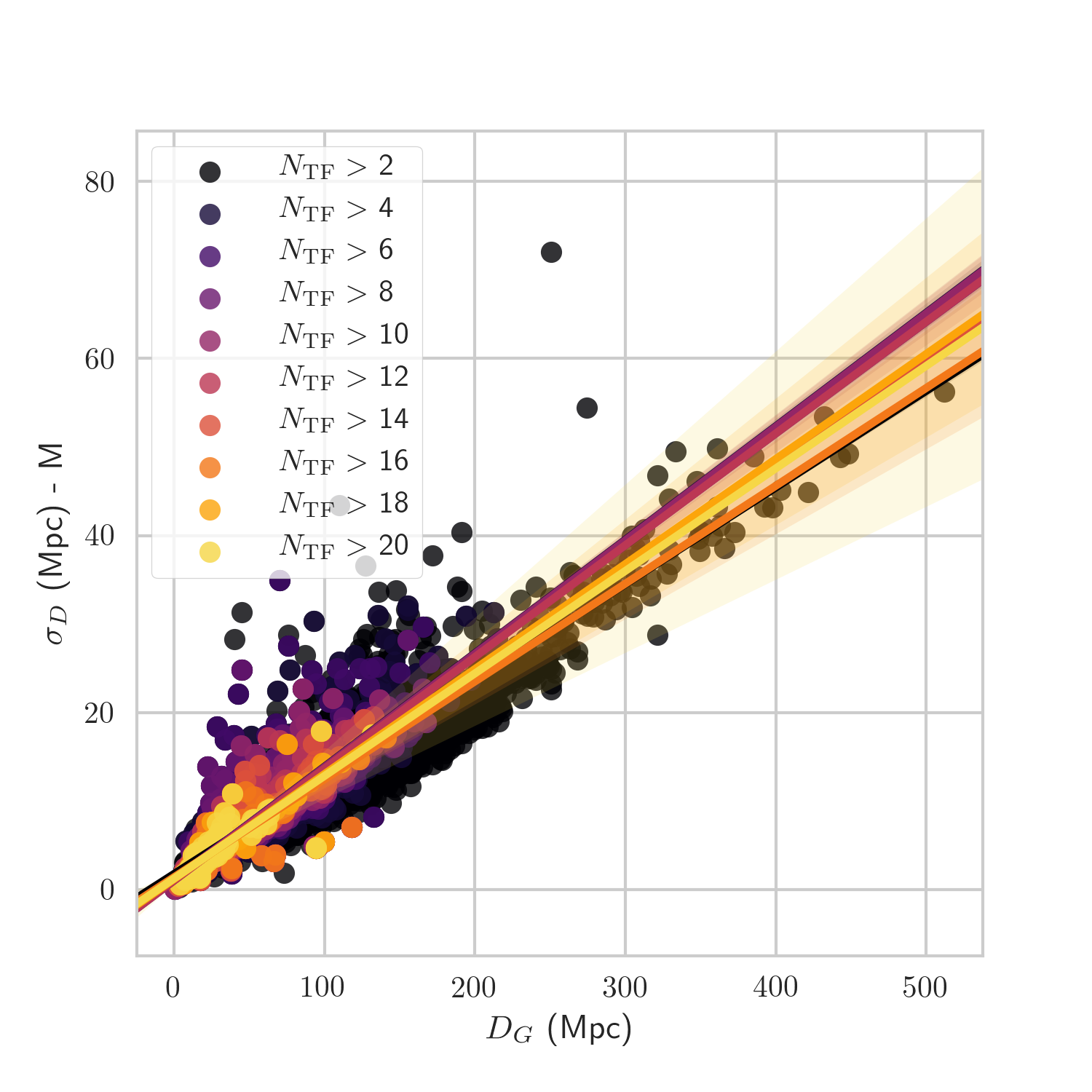}
    \caption{Estimated extragalactic distance errors vs. median extragalactic distance using the Tully-Fisher method for distance determination in NED-D, according to the H error model (left) and the M error model (right) showing linear regressions and confidence intervals computed using the \texttt{seaborn.regplot} Python function.}
    \label{fig:relerr}
\end{figure*}

Fig.~\ref{fig:hqp-qm} shows that the Q and P methods under-predict distance errors for galaxies with more than 5 TF distance measurements. On the other hand, method Q under-predicts distance errors with respect to the M method, which again shows a tighter linear correlation due to the robustness of the M measure. However, the scale of H and M errors (relative errors) does not depend strongly on the limiting number of measurements for $N_\mathrm{TF}>3$, as Fig.~\ref{fig:relerr} shows. \\

\begin{figure*}
	\includegraphics[scale=0.69]{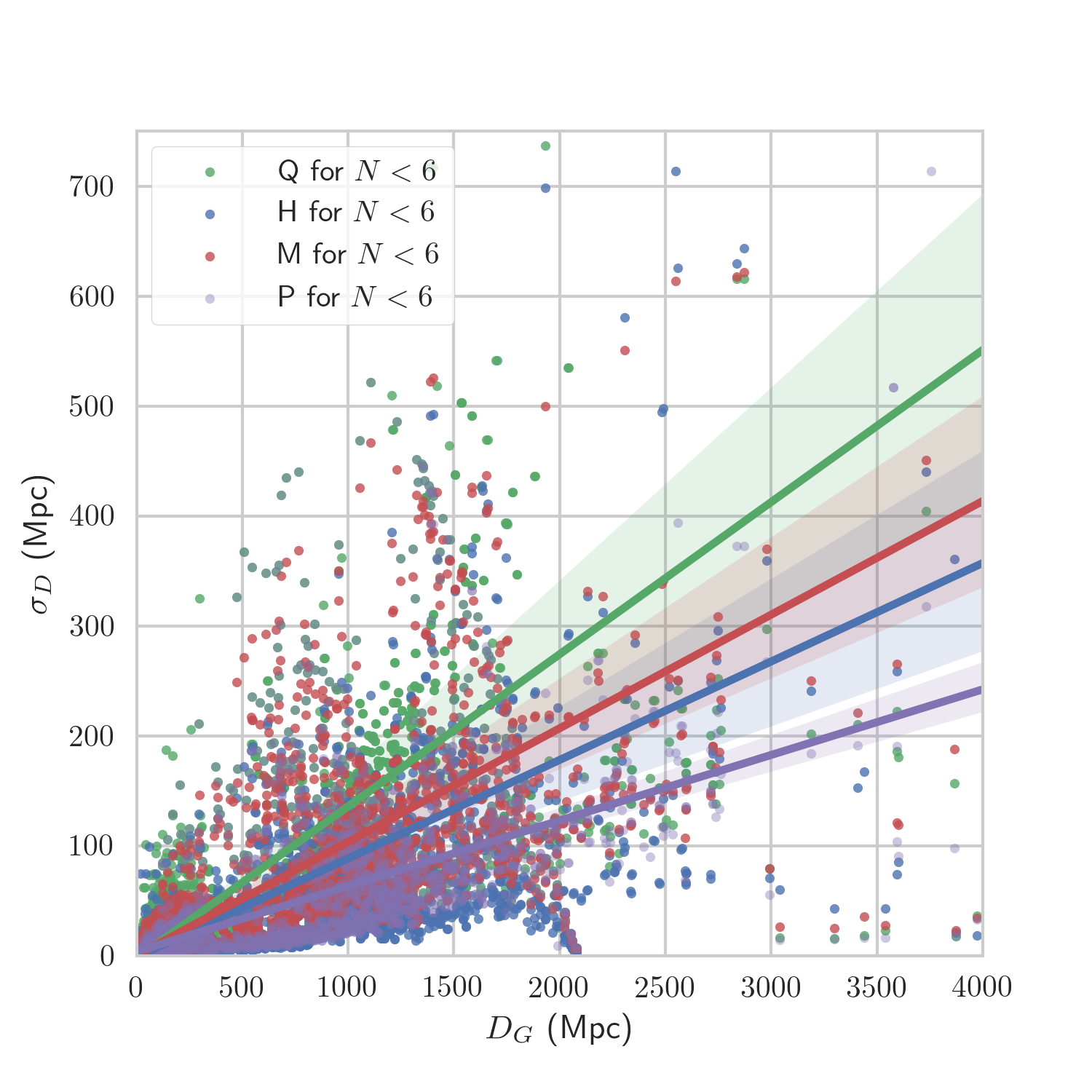}
	\includegraphics[scale=0.69]{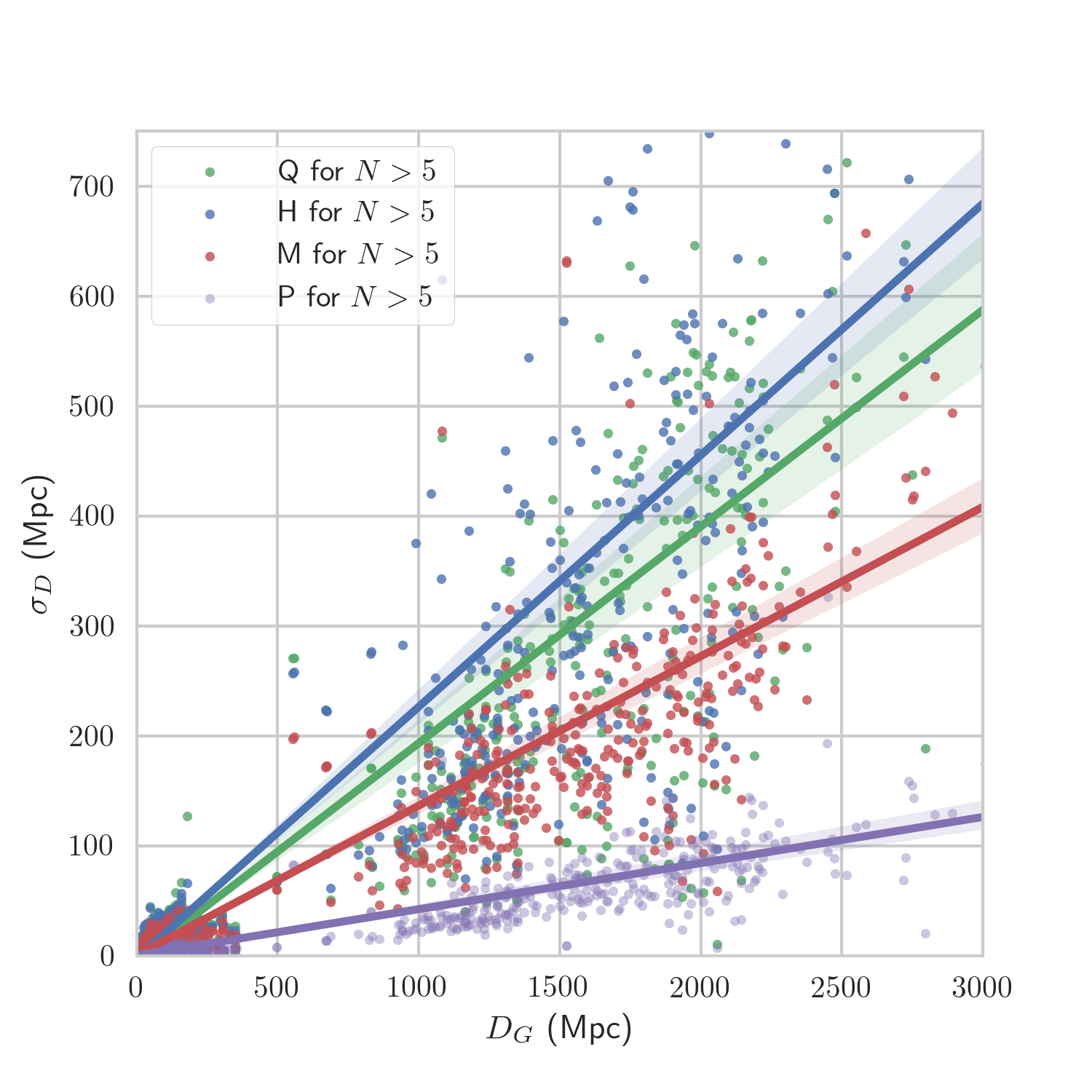}
    \caption{Estimated extragalactic distance errors vs. median extragalactic distance for galaxies with $N<6$ (left) and $N>5$ (right)  redshift-independent distance measurements in NED-D according to the H, M, Q, P error models  (explained in the text), showing linear regressions and confidence intervals computed using the \texttt{seaborn.regplot} Python function.}
    \label{fig:NED}
\end{figure*}

The general correlation between distance and distance error (Figs.~\ref{fig:NED} and \ref{fig:hqp-qm}) means that there is a strong systematic component in the variance of $P(D_G)$, which is expected from the conversion of distance modulus to metric distance. To improve visualization, only errors for galaxies with more than 5 TF distance measurements are shown in Figs.~\ref{fig:hqp-qm}, \ref{fig:drawsee}, and \ref{fig:predl1}.\\


Given that each $\sigma_D$ calculated using the H and M methods is obtained from many realizations from the distribution of extragalactic distances, it is also possible to calculate its variance as the half-distance between the 84th and 16th percentile of bootstrap $\sigma_D$ realizations. Fig.~\ref{fig:drawsee} (left) shows that the variance of the estimated error is proportional to the error for the H and M methods. This will be relevant in Section~\ref{sec:predbay} when we construct a predictive model for non-reported errors.

\section{Predictive Bayesian Models for TF missing errors}
\label{sec:predbay} 

In the multi-measurement catalogs considered here, we observe that the scatter of reported distance measurements and reported individual measurement errors do not match in most cases. This situation happens because there are hidden systematic sources intrisic to each method of distance estimation. These systematics can not be removed, but they can be margizalized over in order to estimate the true variance of a distance estimation method based on multiple measurements. The central limit theorem indicates that as the number of measurements increases, the behavior of the distance errors should settle toward being normally distributed. Thus, if a correlation trend between distance measurements and estimated errors can be explained from a Bayesian viewpoint, then it should be possible to use a Bayesian model to predict missing distance errors for a distance determination method, given enough data. Since more measurements can increase our knowledge of systematic uncertainties in distance measurements, the way we explore and validate our Bayesian models is based on partitioning our data by choosing different lower thresholds for $N$, the number of measurements per galaxy.\\

As seen in Fig.~\ref{fig:hqp-qm}, TF distance errors estimated using the robust methods H and M grow in a roughly linear fashion with distance and seem to be randomly distributed around this trend line. For this reason we try out simple linear and quadratic Bayesian models in order to be able to predict the value of missing distance errors. For this, we use the \texttt{emcee} affine invariant Markov Chain Monte Carlo (MCMC) ensemble sampler \citep{emcee}. Recently, \texttt{emcee} has been proved to be useful in obtaining probabilistic estimations for photometric redshifts \citet{photred1,photred2}. Since we want to be able to predict non-reported errors, our model selection is based on posterior predictive checks, i.e. we rely on models that can create synthetic datasets similar to the original dataset \citep{gelmanppd}. This allows us to reproduce the original variance of the error (Fig.~\ref{fig:drawsee}, left). Many Bayesian analyses often do not use posterior predictive checks, like in the work of \citet{propprob2018} and \citet{bayesh}, where they used \texttt{emcee} for posterior sampling, and instead using Bayesian and Akaike Information Criteria along with Bayes factors for model assessment, but without attempting to reproduce the original variance of the data. This is also the case in other Bayesian tools like Linmix  \citep{gmastro}, which is widely used in astronomy for approximating unobserved data.  \\

First we assume that for any galaxy $j$ the distance error $\sigma_{Dj}$ is a random normal variable, with variance $\sigma_{\sigma j}$ and mean $\hat{\sigma}_{Dj}$, 
\begin{equation}
P(\sigma_{Dj}|\hat{\sigma}_{Dj},\sigma_{\sigma j})=\mathcal{N}(\hat{\sigma}_{Dj},\sigma_{\sigma j}^2)\ .
	\label{eq:prob}
\end{equation}
Our likelihood function is the joint probability that each of the $\sigma_D=\{\sigma_{Dj}\}$ in the original dataset of $m$ galaxies is generated by the above probability,  
\begin{equation}
 P(\sigma_{D}|\hat{\sigma}_{D},\sigma_{\sigma})=\prod_j^mP(\sigma_{Dj}|\hat{\sigma}_{Dj},\sigma_{\sigma j})
\end{equation}
We want to test the hypothesis mentioned above that all errors and their variances $(\hat{\sigma}_D=\{\hat{\sigma}_{Dj}\},\ \sigma_\sigma=\{\sigma_{\sigma j}\})$ can be estimated from a single model depending on the extragalactic distances $D_G=\{D_{Gj}\}$ and a set of distance-independent parameters $\pmb{\theta}$. Thus the likelihood can be expressed as,
\[P(\sigma_D|D_G,\pmb{\theta})=\prod_j^mP(\sigma_{Dj}|D_{Gj},\pmb{\theta})\ .\]
Following Bayes' theorem we can compute the posterior probability up to a constant,
\begin{equation}
P(\pmb{\theta}|D_G,\sigma_D)\propto P(\pmb{\theta})P(\sigma_D|D_G,\pmb{\theta})\ .
	\label{eq:ppd}
\end{equation}
Due to the simplicity of the models used here, we will only use reasoably conservative (flat) priors on all model parameters, which are described in the next subsection.\\

A common feature across our models is that $\sigma_\sigma=f\hat{\sigma}_D$, where the error variance scale factor $f$ is one of the parameters in $\pmb{\theta}$. This model choice is supported by Fig.~\ref{fig:drawsee} (left), which shows a roughly linear correlation between estimated errors and their variances. On the other hand, our models will differ by the proposed functional forms of $\hat{\sigma}_D(D_G,\pmb{\theta})$.\\

We obtain computationally credible samplings of the posterior probability (equation \ref{eqn:ppd})by removing the burn-in steps of the random walk according to the autocorrelation time. We can then create synthetic datasets by drawing a parameter sample $\pmb{\theta}_k$ from the posterior and using it to draw from the likelihood to create a new dataset, i.e. drawing new $\sigma_{Dj}$ from the probability distribution for all galaxies in the original dataset using equation~\ref{eq:prob}. We then assess the validity of the model by comparing synthetic data with the observed (i.e. original) data. This comparison is done by using a discrepancy measure $\mathcal{D}(\sigma_D|\pmb{\theta}_k)$ between data and model-derived expected values for the same data $e=\{e_j(\pmb{\theta}_k)\}$, where $\theta_k$ is drawn from the posterior distribution and $\sigma_D$ can be the observed errors or the model-generated synthetic errors. The discrepancy can be calculated using a statistic like $\chi^2$ \citep{chi2ms,otherdisc}, but here we will work with the Freeman-Tukey discrepancy since it is weight independent \citep{bishopft,brooks}, 
\[\mathcal{D}(\sigma_D|\pmb{\theta}_k)=\sum_j^m(\sqrt{\sigma_{Dj}\vphantom{e_j(\pmb{\theta}_k)}}-\sqrt{e_j(\pmb{\theta}_k)})^2\]
For each parameter draw $k$, it is possible to compare the simulated discrepancy with the observed discrepancy. If the model is representative of the data, then for many parameter draws, the simulated and observed discrepancies should be similar. We can then calculate a Bayesian ``$p$-value'' as the ratio of ``draws when the observed discrepancies are larger than the synthetic discrepancies'' to ``total draws''. If this Bayesian $p$-value is too close to 0 or to 1 we can reject the model, otherwise it is generating synthetic data that is similar to the original data. This is better visualized using a discrepancy plot, where for each draw $k$, a synthetic discrepancy is paired with its corresponding observed discrepancy. If the discrepancy points are roughly equally distributed about the $\mathcal{D}_\mathrm{obs}=\mathcal{D}_\mathrm{sim}$ line, then we cannot reject the model. As mentioned above, we expect that galaxies with the largest number of measurements are sampling more completely the ``true'' distribution of the distance. Therefore we need to find the minimum number of measurements per galaxy for which the Bayesian $p$-value shows an agreement between on the partitioned dataset and the model predictions.

\subsection{Bayesian Quadrature Model}
\label{sec:bqm} 

\begin{figure*}
	\includegraphics[scale=0.69]{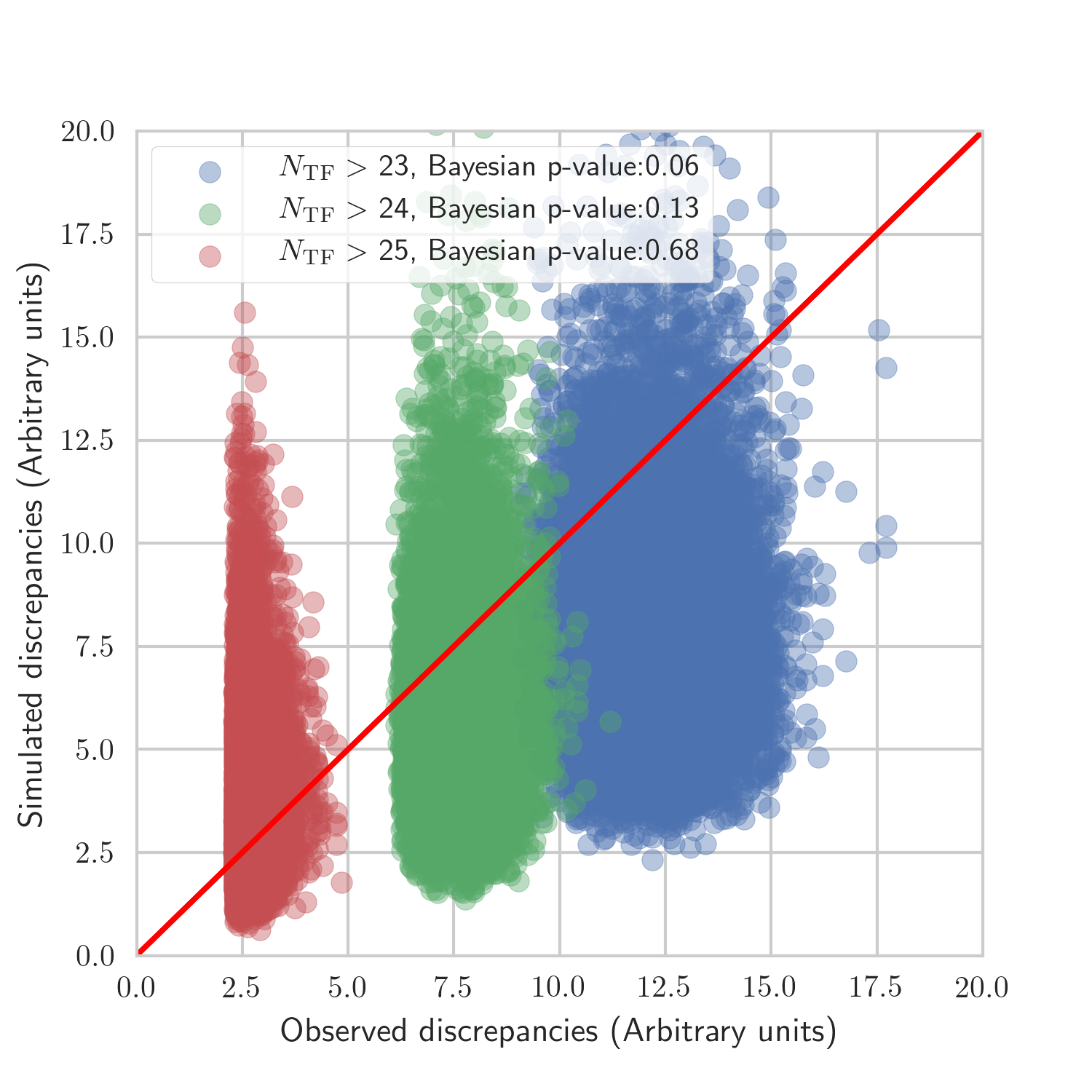}
	\includegraphics[scale=0.69]{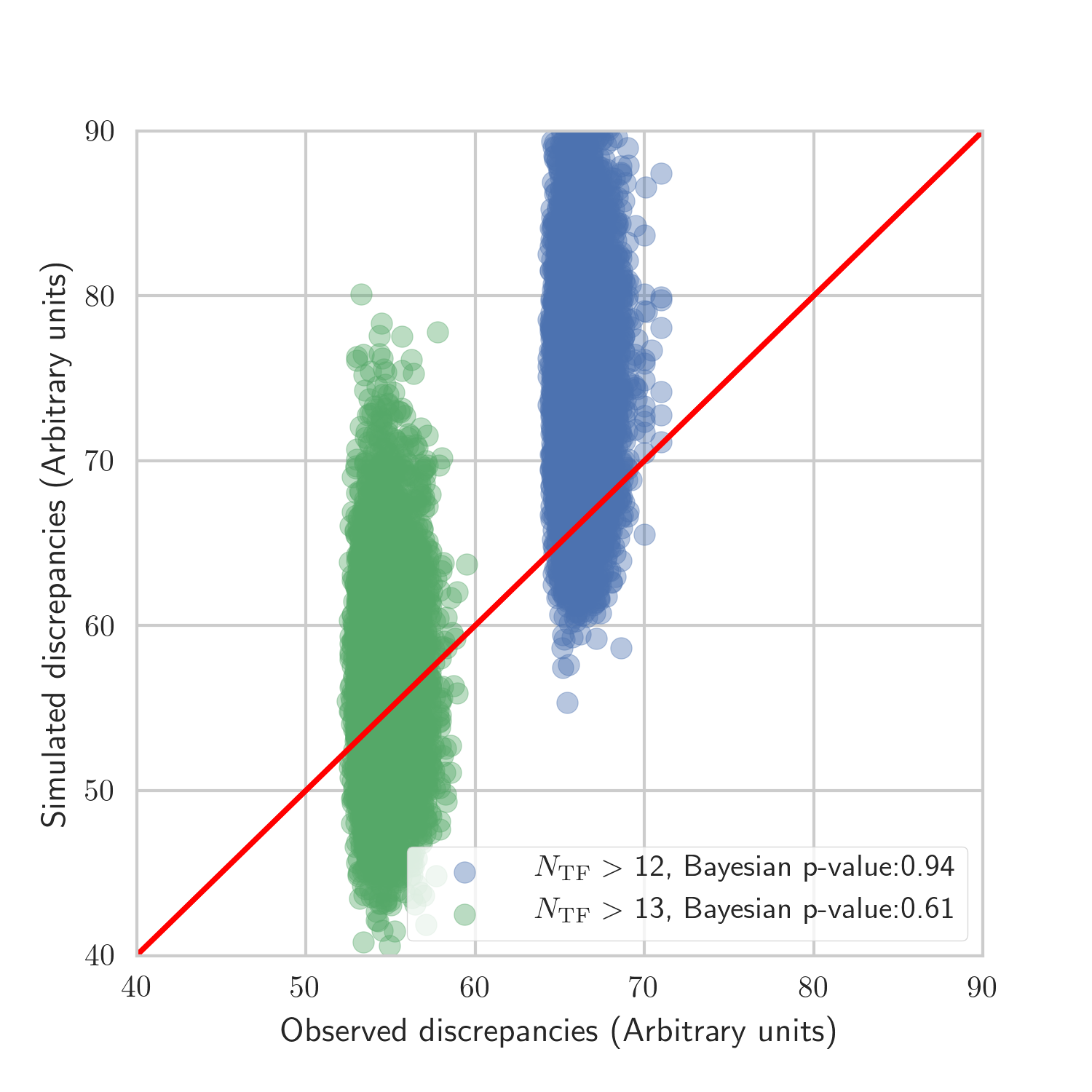}
    \caption{Discrepancy plot for the Bayesian quadrature model (equation~\ref{eq:bayq}) based on errors estimated using method H for $N_\mathrm{TF}>23,24,25$ (left) and using method M for $N_\mathrm{TF}>12,13$ (right).}
    \label{fig:discq}
\end{figure*}

\begin{figure*}
	\includegraphics[scale=0.69]{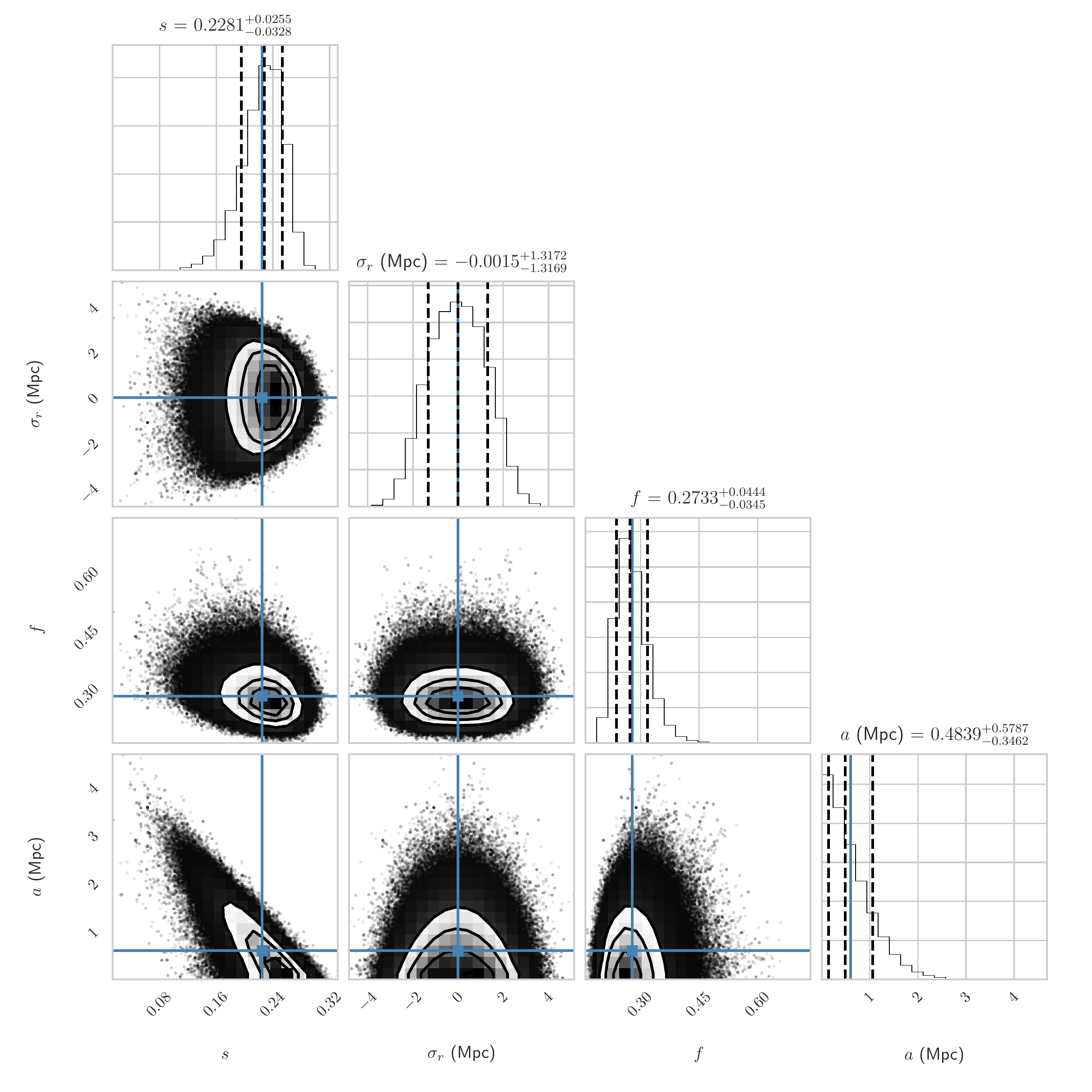}
    \caption{Corner plot showing the \texttt{emcee} sampling of the posterior probability distribution (equation \ref{eq:ppd}) for the quadrature Bayesian model parameters $\pmb{\theta}=(s,\sigma_r,f,a)$ based on errors estimated using method H for galaxies with more than 25 TF distance measurements. The dashed lines indicate the 16th, 50th, and 84th percentile of the marginalized distribution of each parameter (shown at the top of each column), and the blue solid lines indicate the mean. This plot was made using the \texttt{corner} Python module.}
    \label{fig:cornerq}
\end{figure*}

\begin{figure*}
	\includegraphics[scale=0.69]{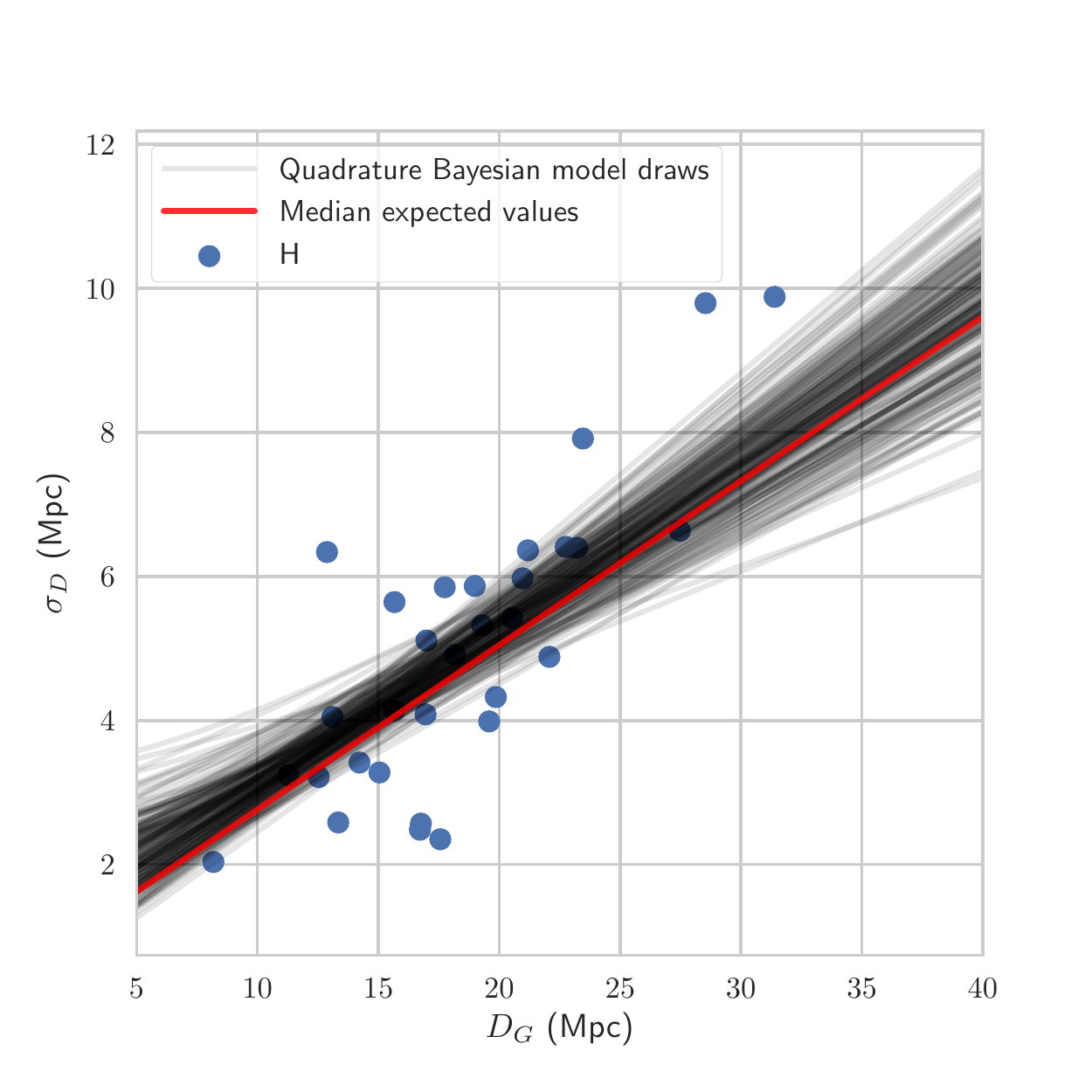}
	\includegraphics[scale=0.69]{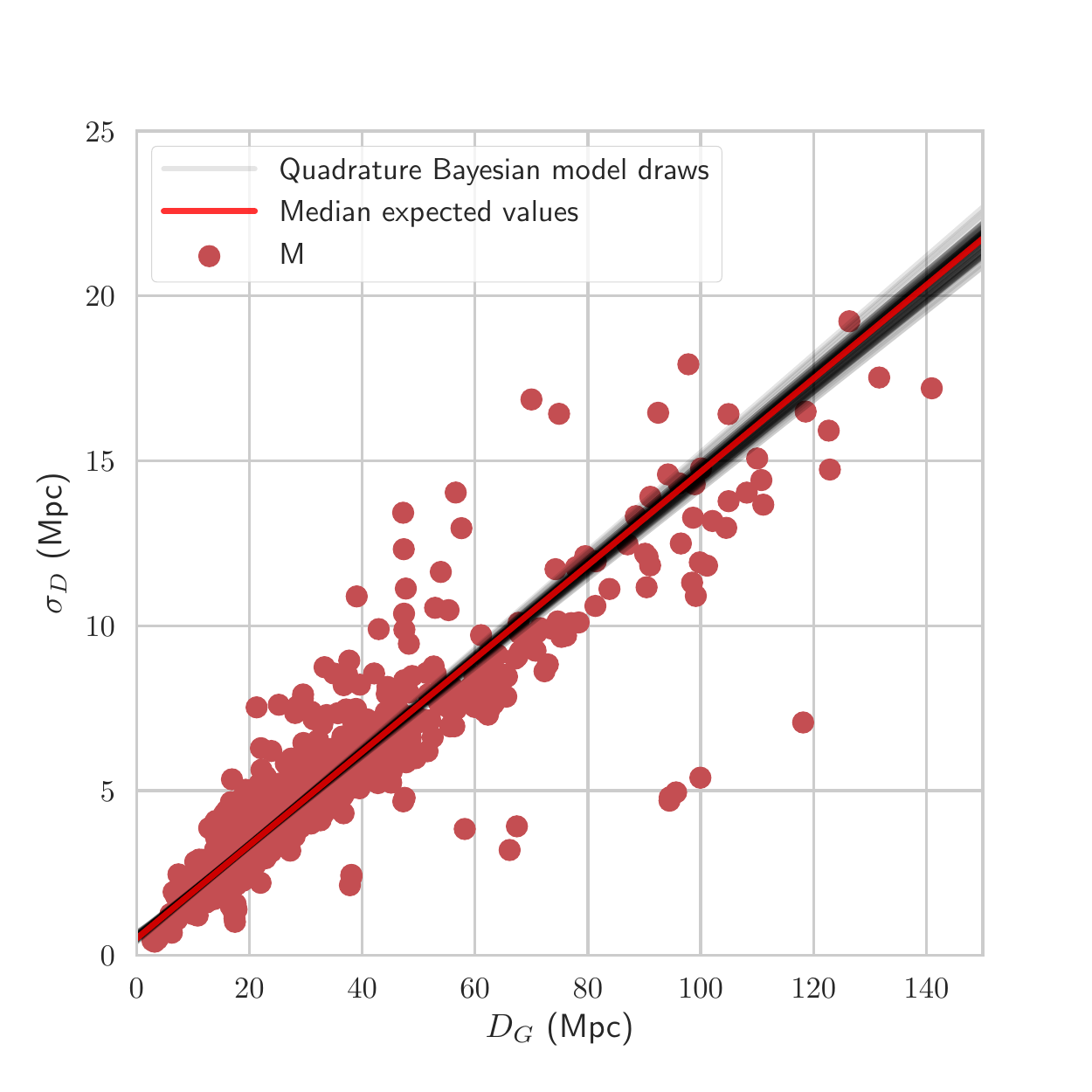}
    \caption{Projection of parameter set samples from the posterior probability distribution of the Bayesian quadrature model onto the  $\sigma_D$ vs. $D_G$ scatter plot for errors estimated using method H for galaxies with more than 25 TF distance measurements (left) and using method M for galaxies with more than 13 TF distance measurements (right).}
    \label{fig:drawsq}
\end{figure*}

Our first model is based on the hypothesis that there are are distinct systematic and random contributions to the distance measurement error, both of which are normally distributed. For this reason they are added in quadrature, 
\begin{equation}
\sigma_D^2=\sigma_s^2+\sigma_r^2\ .
	\label{eq:bayq}
\end{equation}
Here $\sigma_r$ is a random (constant) error and the systematic error is modeled allowing for scale factor ($s$) and zero setting ($a$) errors, i.e.  $\sigma_s=sD+a$, as Fig.~\ref{fig:hqp-qm} suggests.  We set our prior to be symmetrical around $\sigma_r=0$ in order to better visualize its behavior near this point, so
\begin{equation}
P(s,a,\sigma_r,f)\propto\left\{
\begin{aligned}
1,\ \ \ \ &\mathrm{if}\ \ \ 0<s<1\ \mathrm{and}\\
& \ \ \ \ \  0<a<10\ \mathrm{Mpc}\ \mathrm{and}\\
&-10<\sigma_r<10\ \mathrm{Mpc}\ \mathrm{and}\\
& \ \ \ \ \  0<f<1\\
0,\ \ \ \ &\ \mathrm{otherwise.}
\end{aligned}
\right.
	\label{eq:priorq}
\end{equation}
We now use \texttt{emcee} to sample the posterior over the parameter set $\pmb{\theta}=(s,\sigma_r,f,a)$ using 100 walkers and 20000 steps ($\bar{t}_\mathrm{autocorr} \lesssim 90$ steps). According to the discrepancy plot in Fig.~\ref{fig:discq} (left), this model is able to replicate method H errors for the 31 galaxies with $N>25$ measurements (866 measurements in total). The corner plot showing the posterior sampling made by \texttt{emcee} is shown in Fig.~\ref{fig:cornerq}, which shows the 16th, 50th, and 84th percentiles of the marginalized posterior distributions for the systematic scale factor $s$, the random error component $\sigma_r$, the error variance scale factor $f$, and the zero offset systematic error $a$. From the large variance in the marginalized posterior distribution for $\sigma_r$ and $a$, we see that there is a significant degeneracy between those parameters. However, it should be noted that the marginalized posterior distribution of $\sigma_r$ is symmetric around zero (because of its own degeneracy), while the distribution of $a$ can only take positive values. The working distance range and overall fitting of this model is shown in Fig.~\ref{fig:drawsq} (left), where method H errors corresponding to galaxies with more than 25 TF distance measurements are plotted along the expected values $e=\{e_j(\pmb{\theta}_k)\}$ for parameter sets $\pmb{\theta}_k$ drawn from the posterior probability distribution. 
\begin{figure*}
	\includegraphics[scale=0.69]{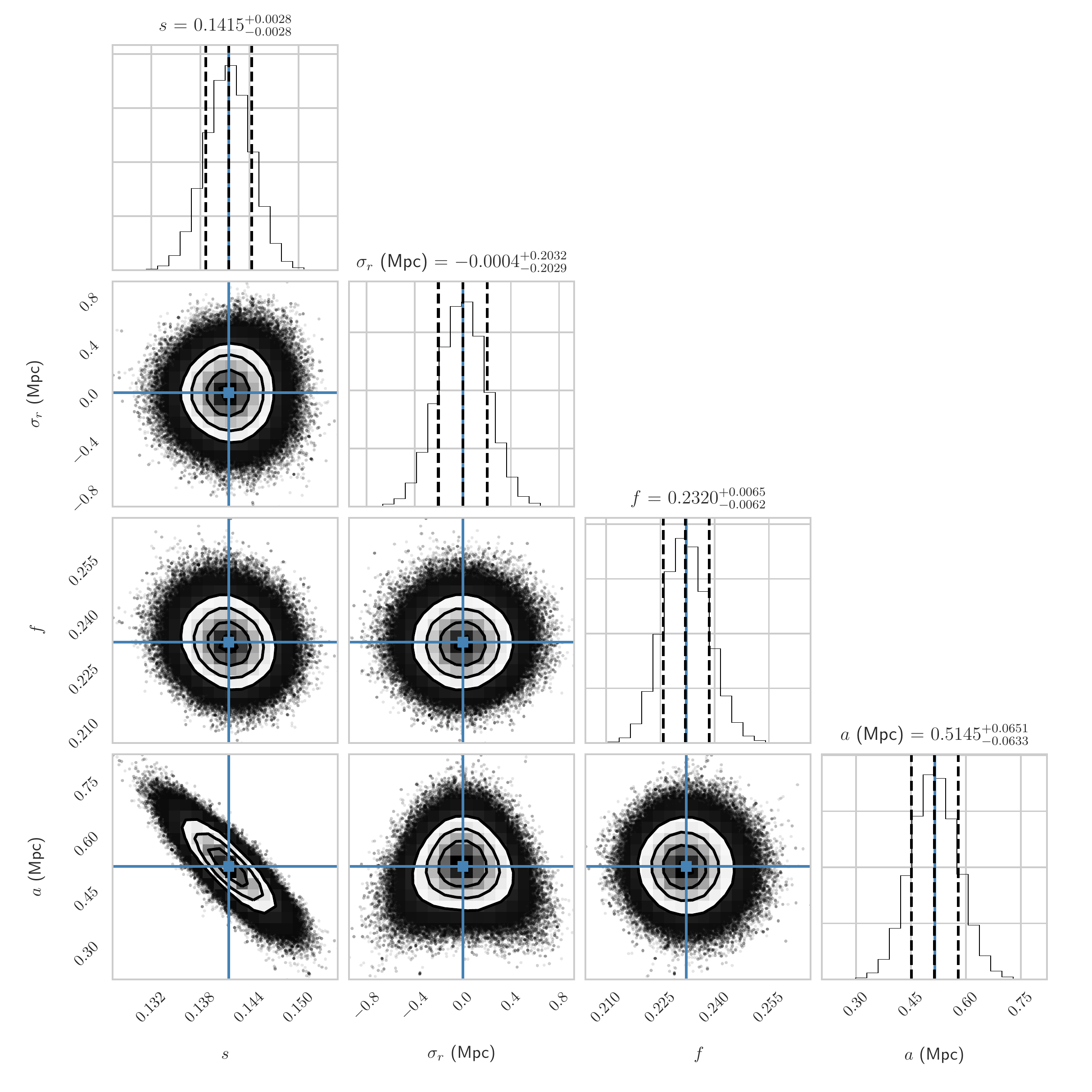}
    \caption{Corner plot showing the \texttt{emcee} sampling of the posterior probability distribution (equation \ref{eq:ppd}) for the quadrature Bayesian model parameters $\pmb{\theta}=(s,\sigma_r,f,a)$ based on errors estimated using method M for galaxies with more than 13 TF distance measurements. The dashed lines indicate the 16th, 50th, and 84th percentile of the marginalized distribution of each parameter (shown at the top of each column), and the blue solid lines indicate the mean. This plot was made using the \texttt{corner} Python module.}
    \label{fig:cornerq2}
\end{figure*}
Now we sample the posterior distribution for the Bayesian quadrature model with method M errors using \texttt{emcee} with 100 walkers and 20000 steps s ($\bar{t}_\mathrm{autocorr} \lesssim 50$ steps). The discrepancy plot for method M errors in Fig.~\ref{fig:discq} (right) shows that the quadrature model also replicates method M errors, but for the 732 galaxies with more than 13 measurements (13054 measurements in total). Fig.~\ref{fig:cornerq2}, shows that values for the random error component $\sigma_r$ are so low that the model draws are almost indistinguishable from straight lines in Fig~\ref{fig:drawsq} (right). Additionally, and just as for the quadrature model for H errors above, the symmetry of the marginalized posterior distribution of $\sigma_r$ leads us to set this parameter to zero in our next (linear) model.

\subsection{Bayesian Linear Model}
\label{sec:blm}
\begin{figure*}
	\includegraphics[scale=0.69]{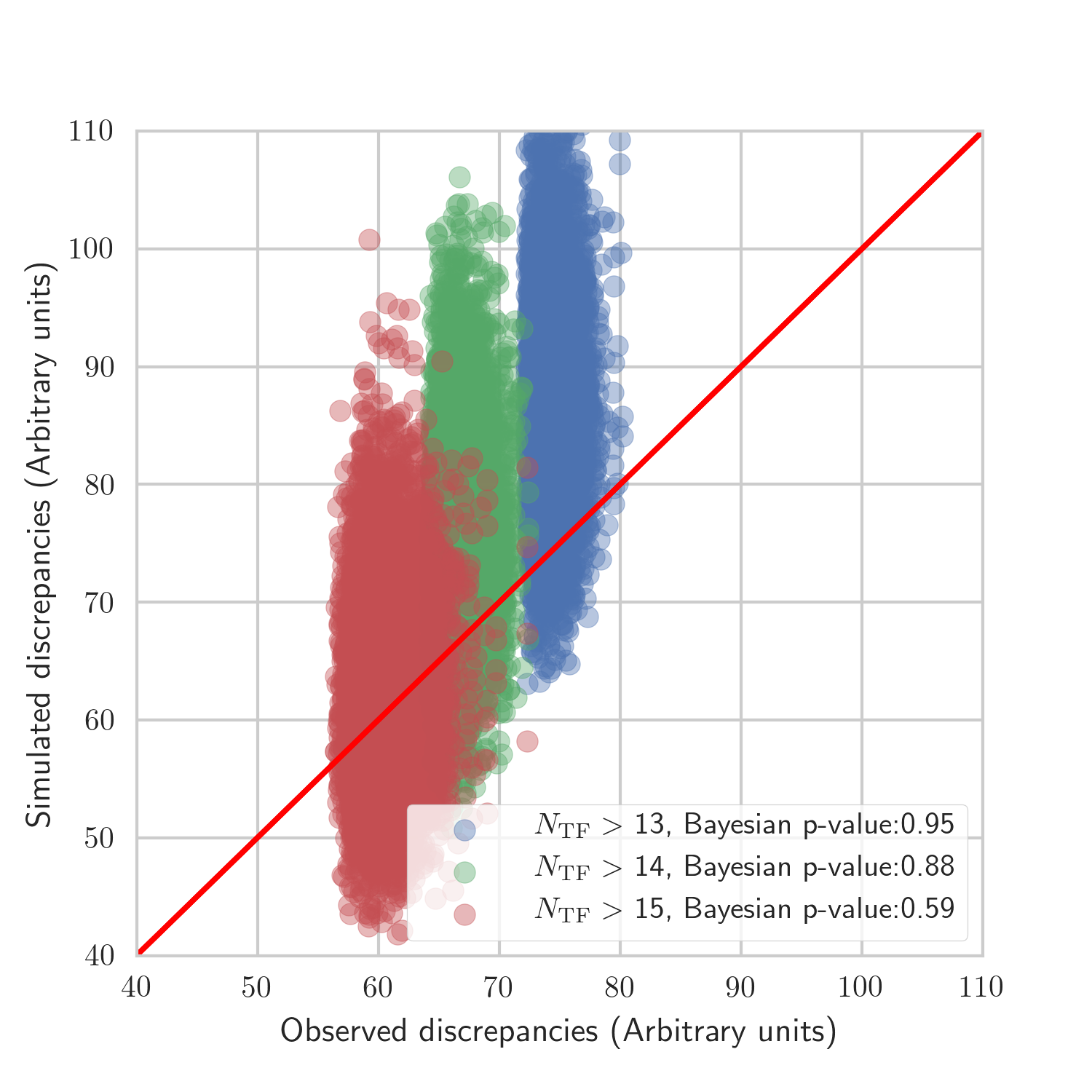}
	\includegraphics[scale=0.69]{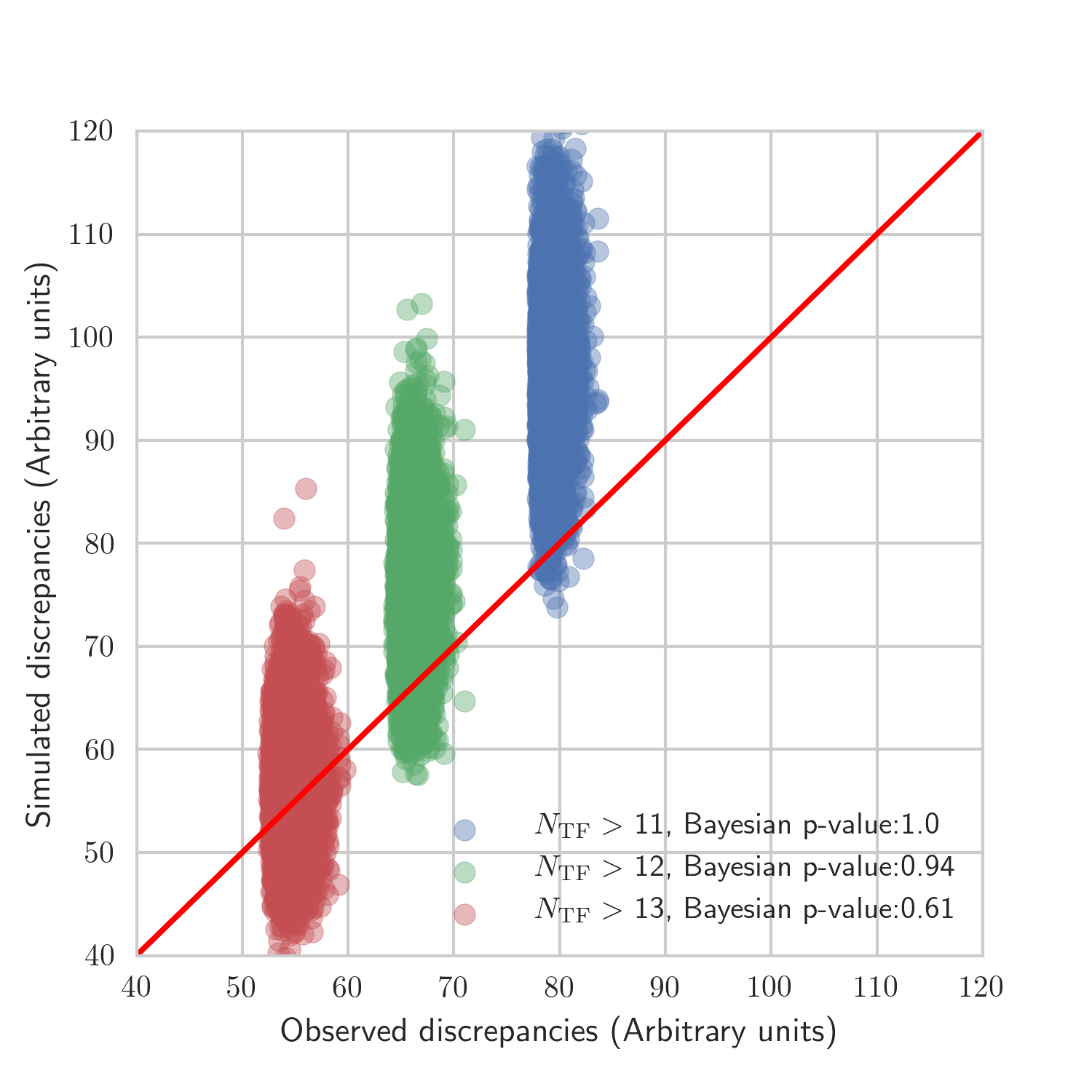}
    \caption{Discrepancy plot for the Bayesian linear model (equation~\ref{eq:bayl}) based on errors estimated using method H for $N_\mathrm{TF}>13,14,15$ (left) and using method M for $N_\mathrm{TF}>11,12,13$ (right). }
    \label{fig:discl}
\end{figure*}

\begin{figure*}
	\includegraphics[scale=0.69]{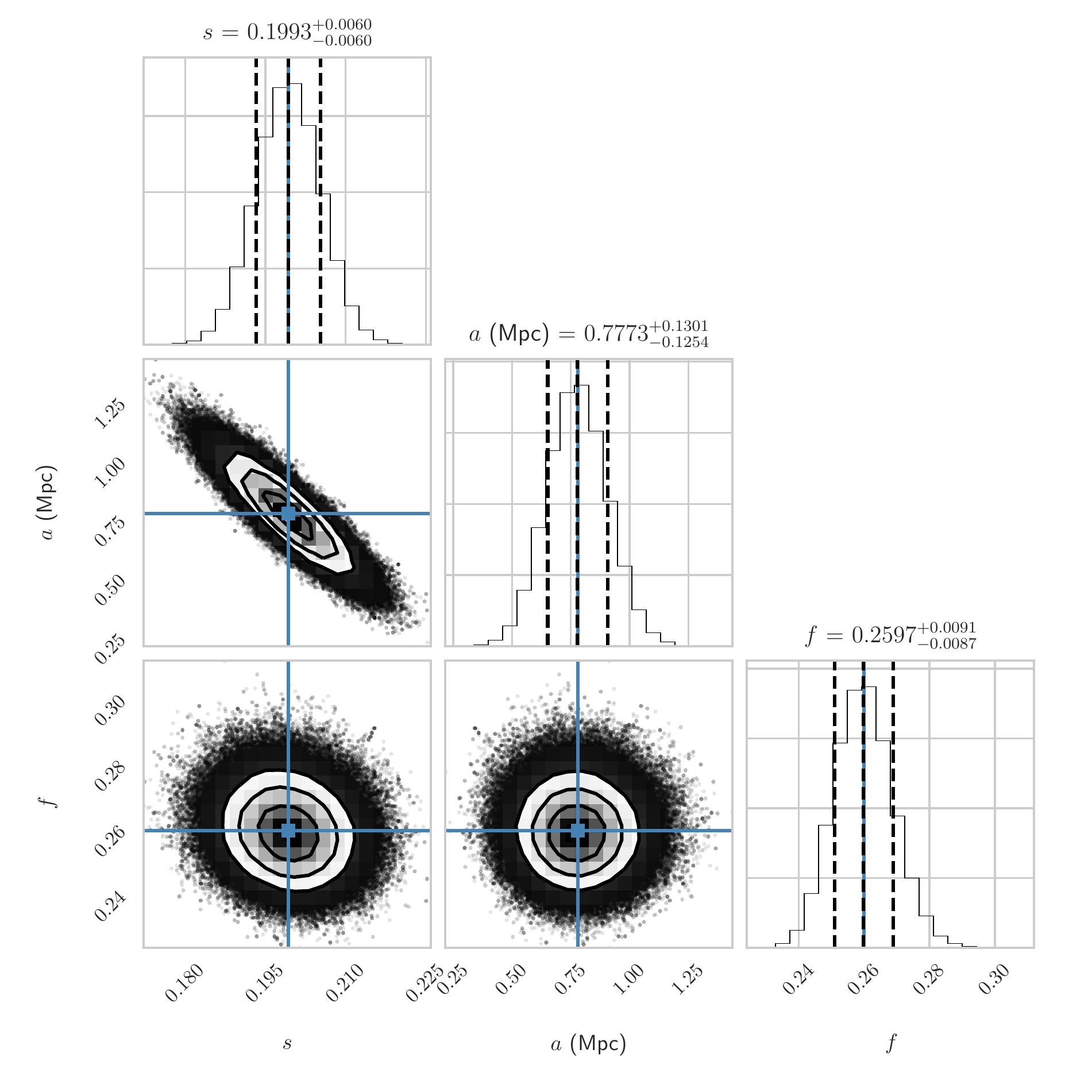}
    \caption{Corner plot showing the \texttt{emcee} sampling of the posterior probability distribution (equation \ref{eq:ppd}) for the linear Bayesian model parameters $\pmb{\theta}=(s,a,f)$ based on errors estimated using method H for galaxies with more than 15 TF distance measurements. The dashed lines indicate the 16th, 50th, and 84th percentile of the marginalized distribution of each parameter (shown at the top of each column), and the blue solid lines indicate the mean. This plot was made using the \texttt{corner} Python module.}
    \label{fig:cornerl}
\end{figure*}

\begin{figure*}
	\includegraphics[scale=0.69]{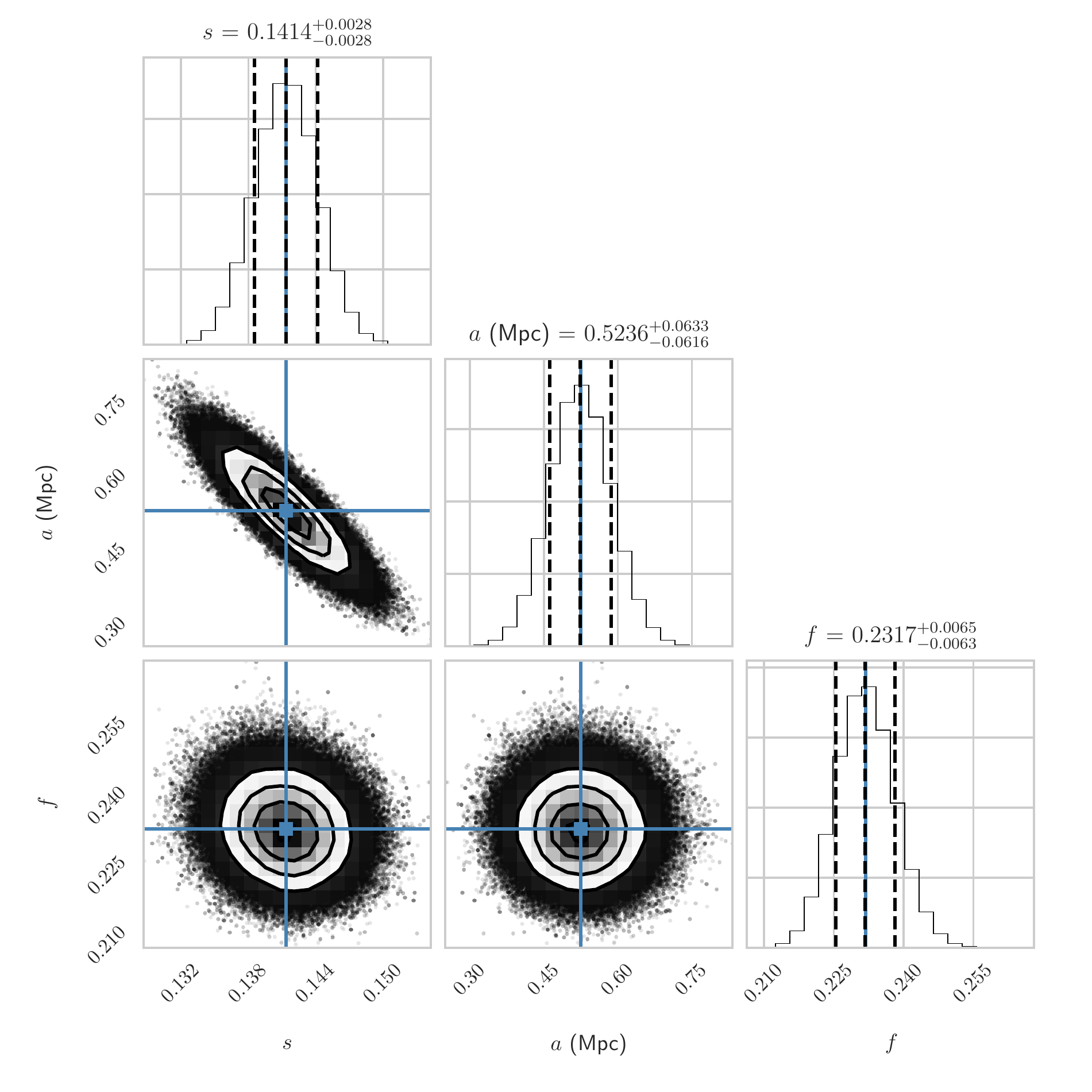}
    \caption{Corner plot showing the \texttt{emcee} sampling of the posterior probability distribution (equation \ref{eq:ppd}) for the linear Bayesian model parameters $\pmb{\theta}=(s,a,f)$ based on errors estimated using method M for galaxies with more than 13 TF distance measurements. The dashed lines indicate the 16th, 50th, and 84th percentile of the marginalized distribution of each parameter (shown at the top of each column), and the blue solid lines indicate the mean. This plot was made using the \texttt{corner} Python module.}
    \label{fig:cornerl2}
\end{figure*}

In Section~\ref{sec:bqm} above we conclude that we can ignore the random error component in equation~\ref{eq:bayq} in order to work with a simpler, numerically stable, linear model that only considers a systematic error with scale factor and zero setting error components,
\begin{equation}
\sigma_D=\sigma_s=sD+a\ .
	\label{eq:bayl}
\end{equation}
We also update our prior considering that the quadratic model yielded lower values for the zero setting error $a$ than previously considered in equation~\ref{eq:priorq},
\begin{equation}
P(s,a,f)\propto\left\{
\begin{aligned}
1,\ \ \ \ &\mathrm{if}\ \ \ 0<s<1\ \mathrm{and}\\
& \ \ \ \ \  0<a<2\ \mathrm{Mpc}\ \mathrm{and}\\
& \ \ \ \ \  0<f<1\\
0,\ \ \ \ &\ \mathrm{otherwise.}
\end{aligned}
\right.
\end{equation}
We use \texttt{emcee} to sample the posterior over $\pmb{\theta}=(s,a,f)$ using 100 walkers and 10000 steps ($\bar{t}_\mathrm{autocorr} < 50$ steps) for the linear Bayesian model applied to H errors. The discrepancy plot (Fig.~\ref{fig:discl}, left) shows a significant improvement over the quadratic model, as it shows an acceptable Bayesian $p$-value for the 477 galaxies with $N>15$ measurements (9361 in total), whereas the quadratic model replicated errors only for galaxies with $N>25$ measurements. Fig.~\ref{fig:cornerl}, shows the 16th, 50th, and 84th percentiles of the marginalized posterior distributions for the systematic scale factor $s$, the error variance scale factor $f$, and the zero offset systematic error $a$ for the linear Bayesian model using H errors for galaxies with more than 15 measurements.
We sample the posterior for the linear model applied to M errors using \texttt{emcee} with 100 walkers and 10000 steps ($\bar{t}_\mathrm{autocorr} < 50$ steps). Fig.~\ref{fig:discl} (right) shows the corresponding discrepancy plot, which does not show a significant improvement of the linear over the quadratic model for M errors, as it also works for galaxies with $N>13$ measurements. This happens because the sampling of the posterior for the quadratic model (Fig.~\ref{fig:cornerq2}) does not show a degeneracy between $\sigma_r$ and $a$, and also because the marginalized posterior distribution for $\sigma r$ is a near-zero-centered distribution with a variance of $\sim0.2$ Mpc. The 16th, 50th, and 84th percentiles of the marginalized posterior distributions for the systematic scale factor $s$, the error variance scale factor $f$, and the zero offset systematic error $a$ according the the linear model for M errors are shown in Fig.~\ref{fig:cornerl2}.

\begin{figure*}
	\includegraphics[scale=0.69]{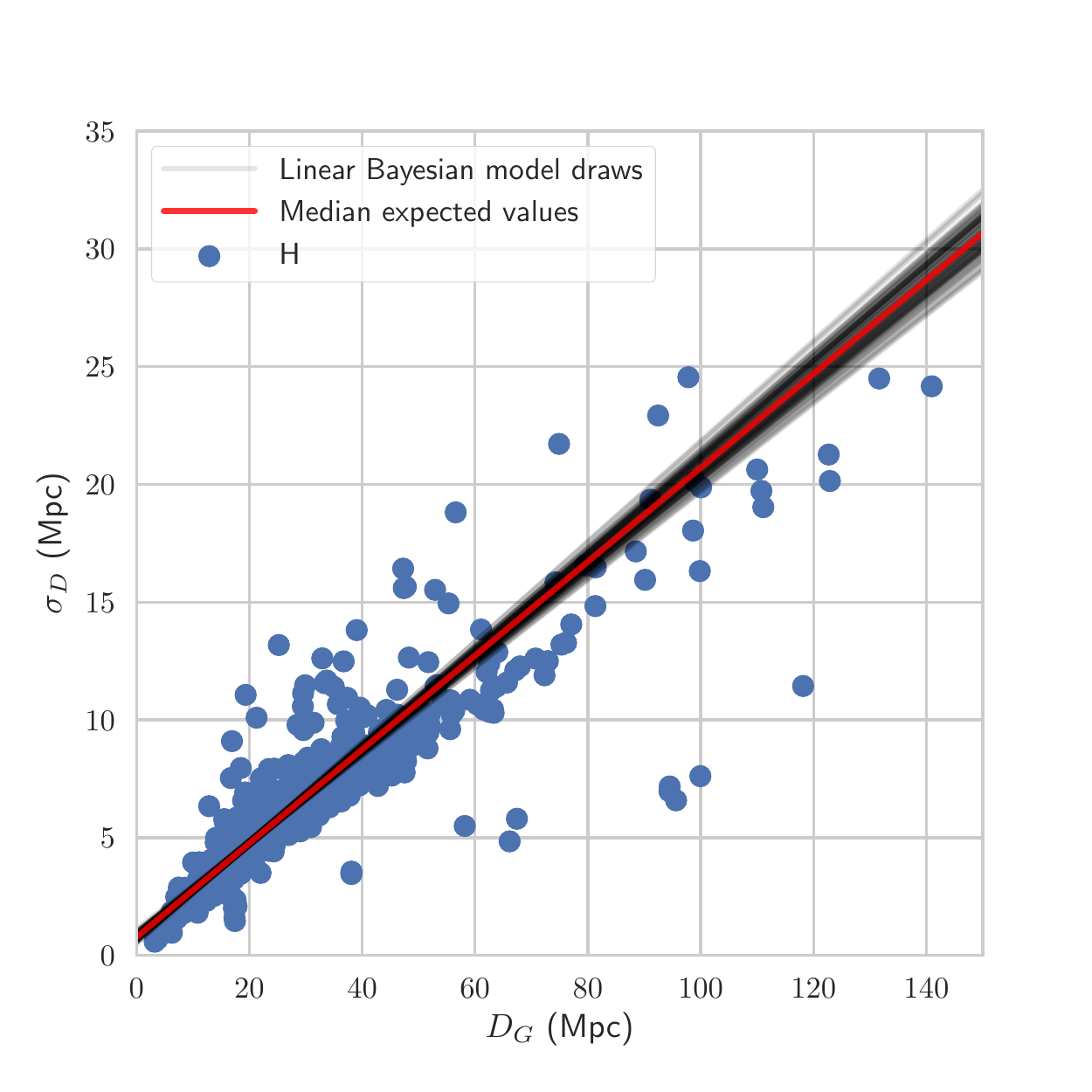}
	\includegraphics[scale=0.69]{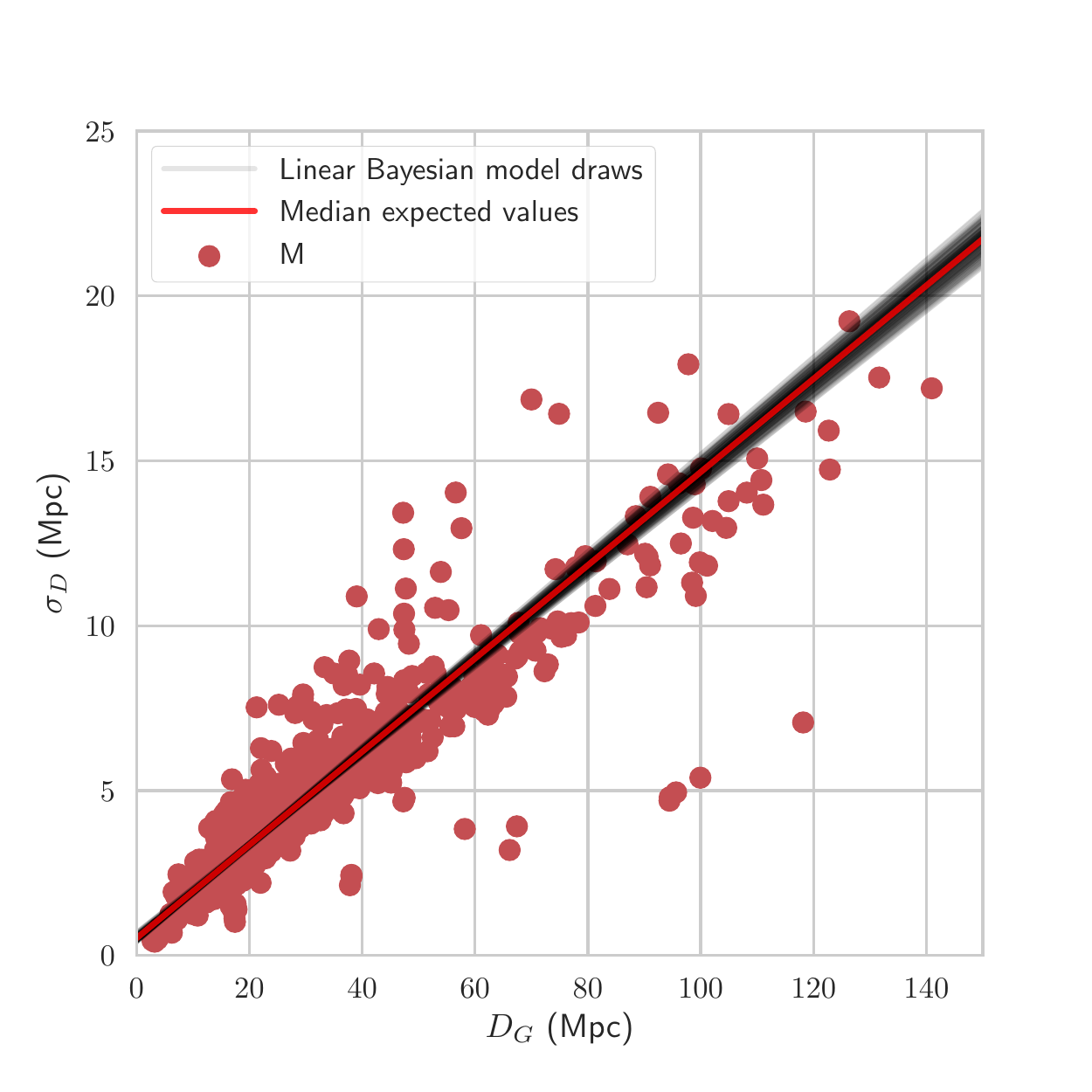}
    \caption{Projection of parameter set samples from the posterior probability distribution of the Bayesian linear model onto the $\sigma_D$ vs. $D_G$ scatter plot for errors estimated using method H for galaxies with more than 15 TF distance measurements (left) and using method M for galaxies with more than 13 TF distance measurements (right).}
    \label{fig:drawsl}
\end{figure*}
\begin{figure*}
	\includegraphics[scale=0.69]{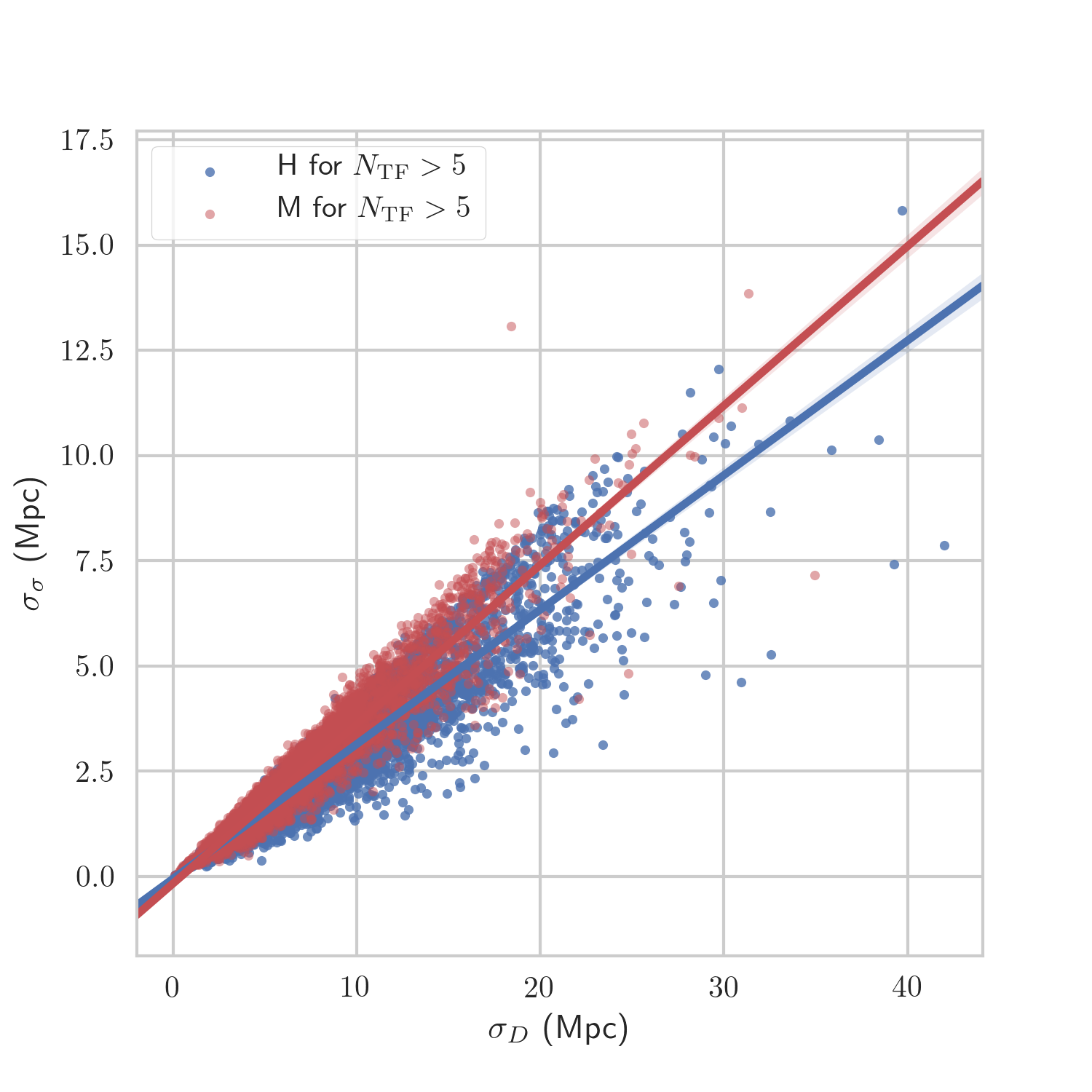}
	\includegraphics[scale=0.69]{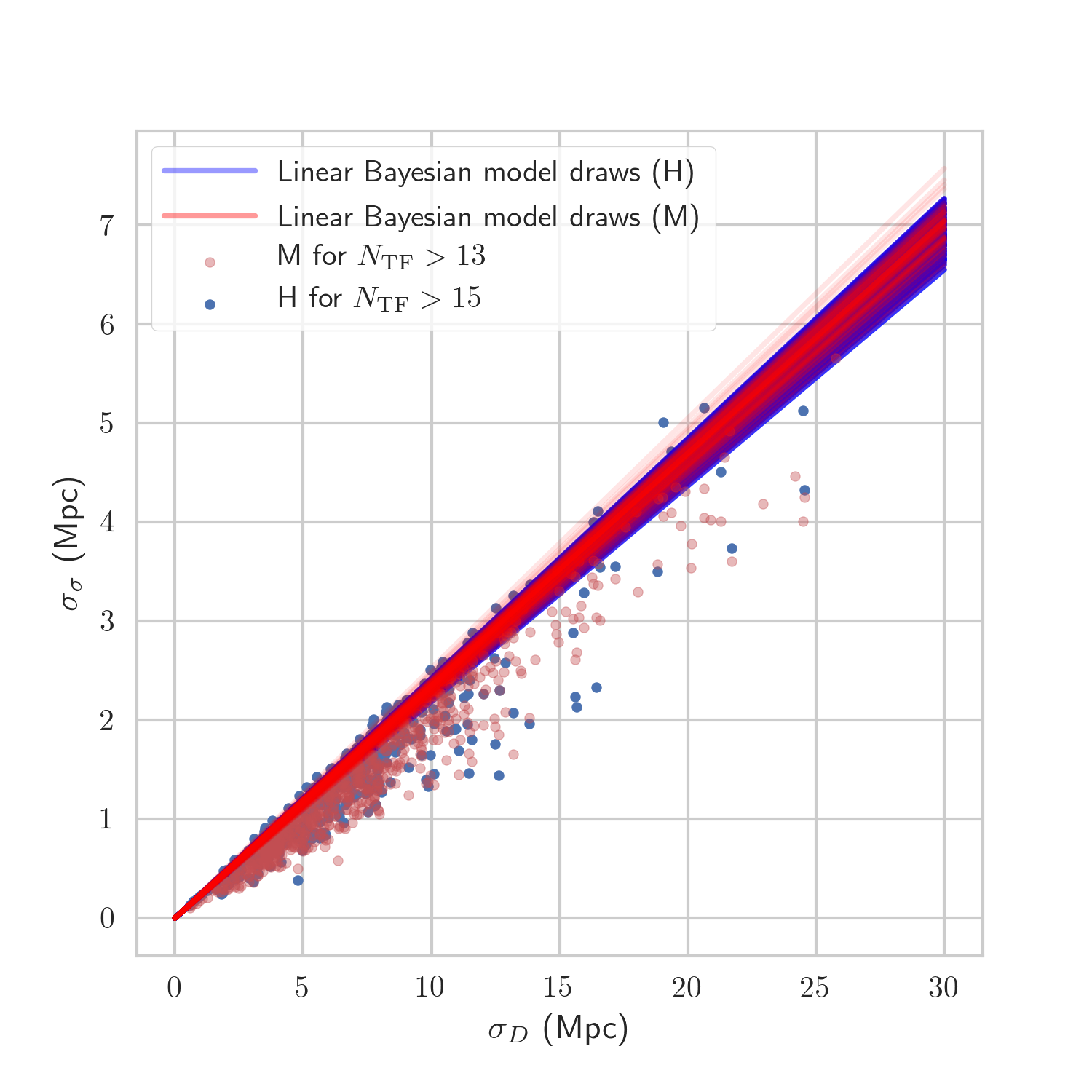}
    \caption{Variance of distance error estimates vs. estimated extragalactic distance errors, showing linear regressions and confidence intervals computed using the \texttt{seaborn.regplot} Python function (left), and showing a projection of parameter set samples from the posterior probability distribution of the Bayesian linear model (right) as determined by the H and M methods.}
    \label{fig:drawsee}
\end{figure*}
\subsection{Predictions for missing errors}
\label{sec:pred} 
Our linear Bayesian model is able to predict the intrinsic variance of TF H and M distance errors in NED-D by considering systematic zero setting and scale factor error components. The lower limit of distance measurements for which the model works for H and M errors, is 15 and 13, respectively\footnote{Our model validation has also worked with the two 2017 versions of the NED-D extragalactic distance catalog, albeit with different thresholds for the number of measurements per galaxy.}. Fig.~\ref{fig:drawsl} shows the linear model draws for H and M errors, for which the working range is approximately $D_G\in[3,140]$ Mpc. We also show in Fig.~\ref{fig:drawsee} that the model draws for $f$ the scale parameter for the variance of $\sigma_D$ fit the bootstrap variance of H and M errors well, which means that our model choice for the variance of the error ($\sigma_\sigma$) was the right one. \\

\begin{figure*}
	\includegraphics[scale=0.69]{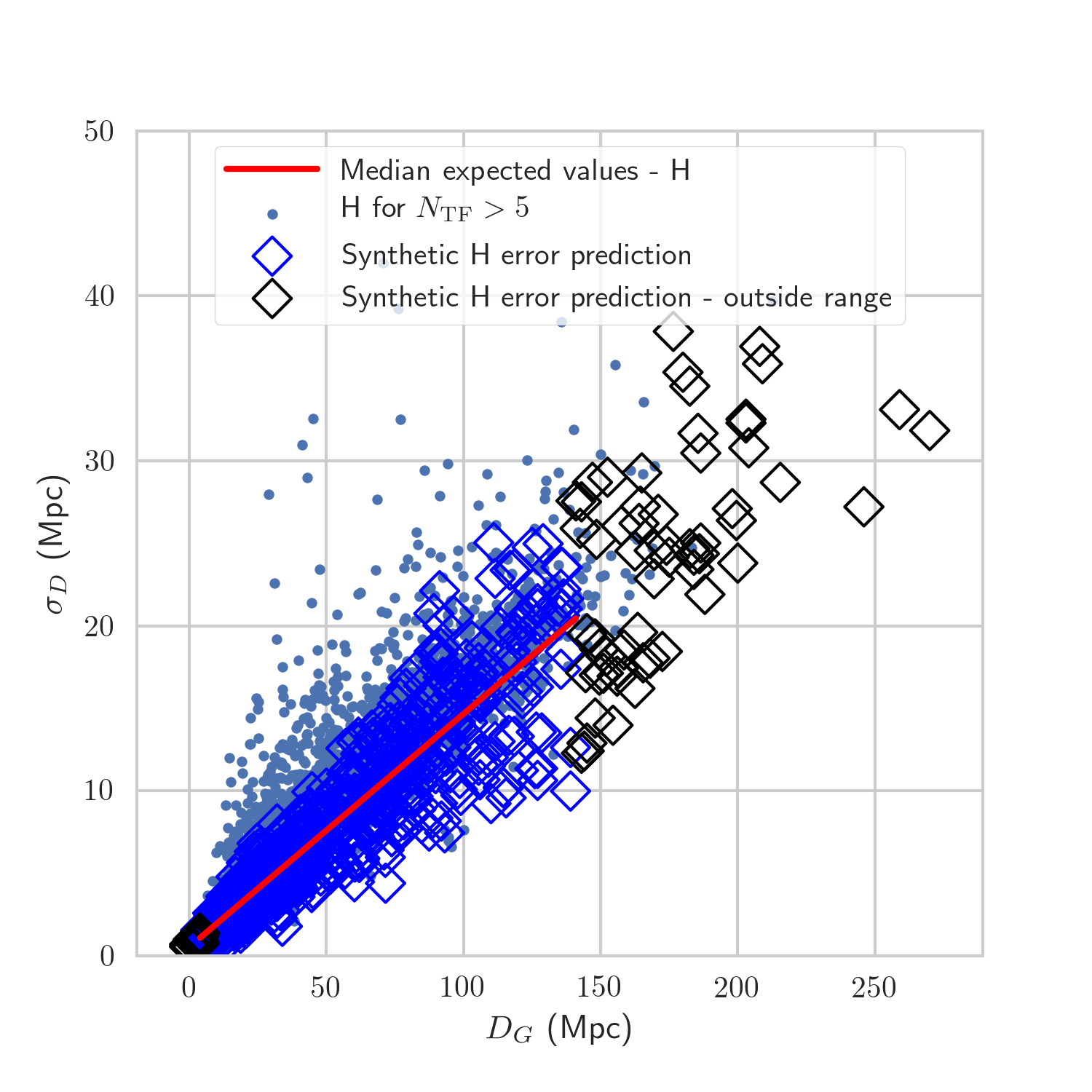}
	\includegraphics[scale=0.69]{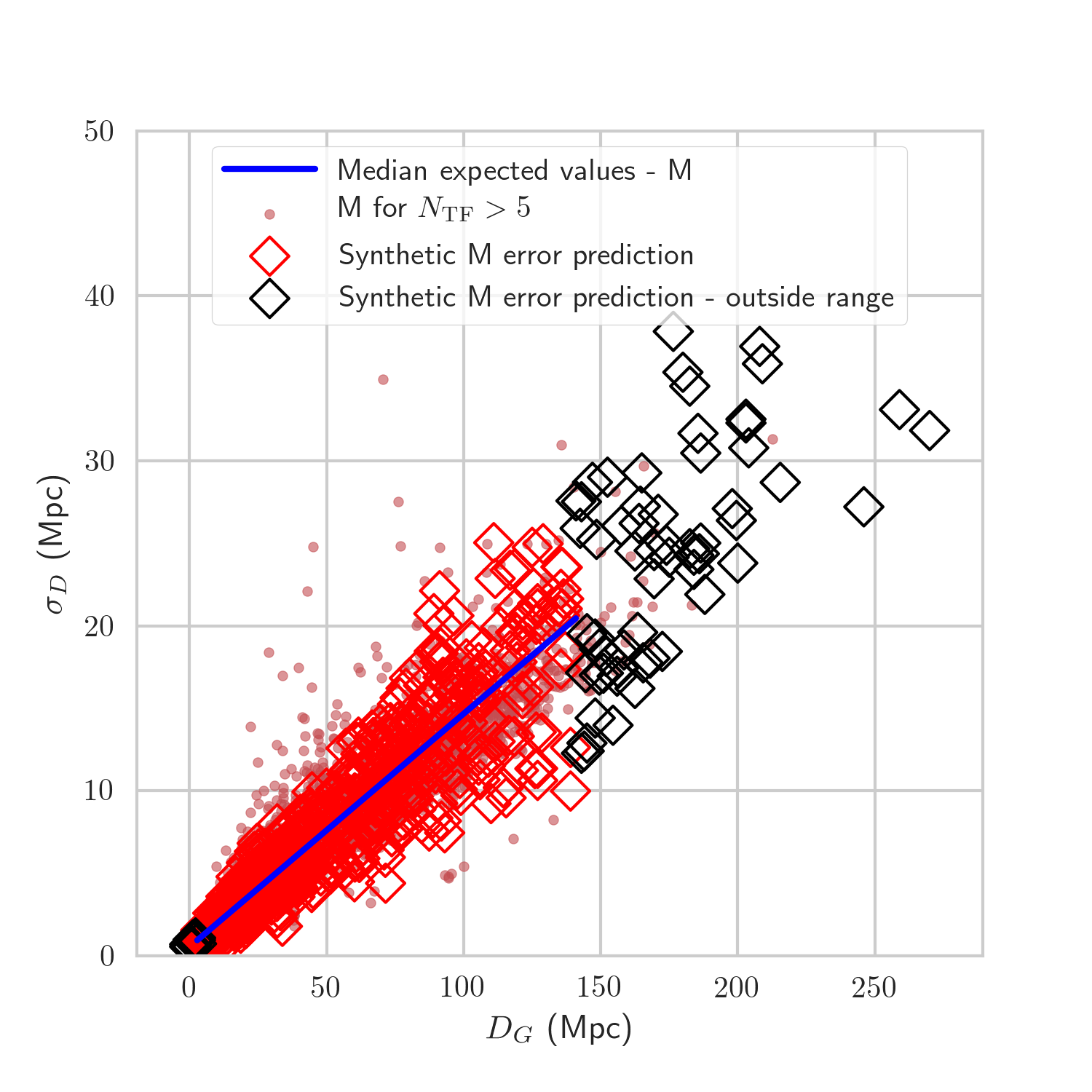}
    \caption{Synthetic H-method (left) and M-Method (right) $\sigma_D$ and their median expected values vs. $D_G$ for the 884 galaxies in NED-D for which no TF distance measurements report an error, generated using the corresponding Bayesian linear model. Predicted errors for galaxies outside of the working distance range of the model are plotted in black. H (left) and M (right) errors for galaxies with more than 5 TF measurements are also plotted for comparison.}
    \label{fig:predl1}
\end{figure*}

Now, galaxies for which the models shown above work are not intrinsically different to other galaxies, as long as they are within the same distance range. Thus, we use the posterior predictive distribution of the linear Bayesian model for predicting H and M errors for the 884 galaxies in NED-D for which all TF measurements lack a reported error. Fig.~\ref{fig:predl1} shows synthetic errors generated from the posterior predictive distribution for the $\sigma_D$ linear model, along with the expected values of $\sigma_D$ using the median of the posterior probability distribution in equation~\ref{eq:ppd}, and the $D_G$ vs. $\sigma_D$ points for galaxies with more than 5 TF distance measurements (for contrast) for methods H and M, respectively. The median expected values are only drawn for points within the predictive range of each model, and synthetic predicted errors for galaxies outside of this range are plotted in black. The distance was calculated using the median of the reported distances whenever there was more than one TF distance measurement.\\

\begin{figure*}
	\includegraphics[scale=0.69]{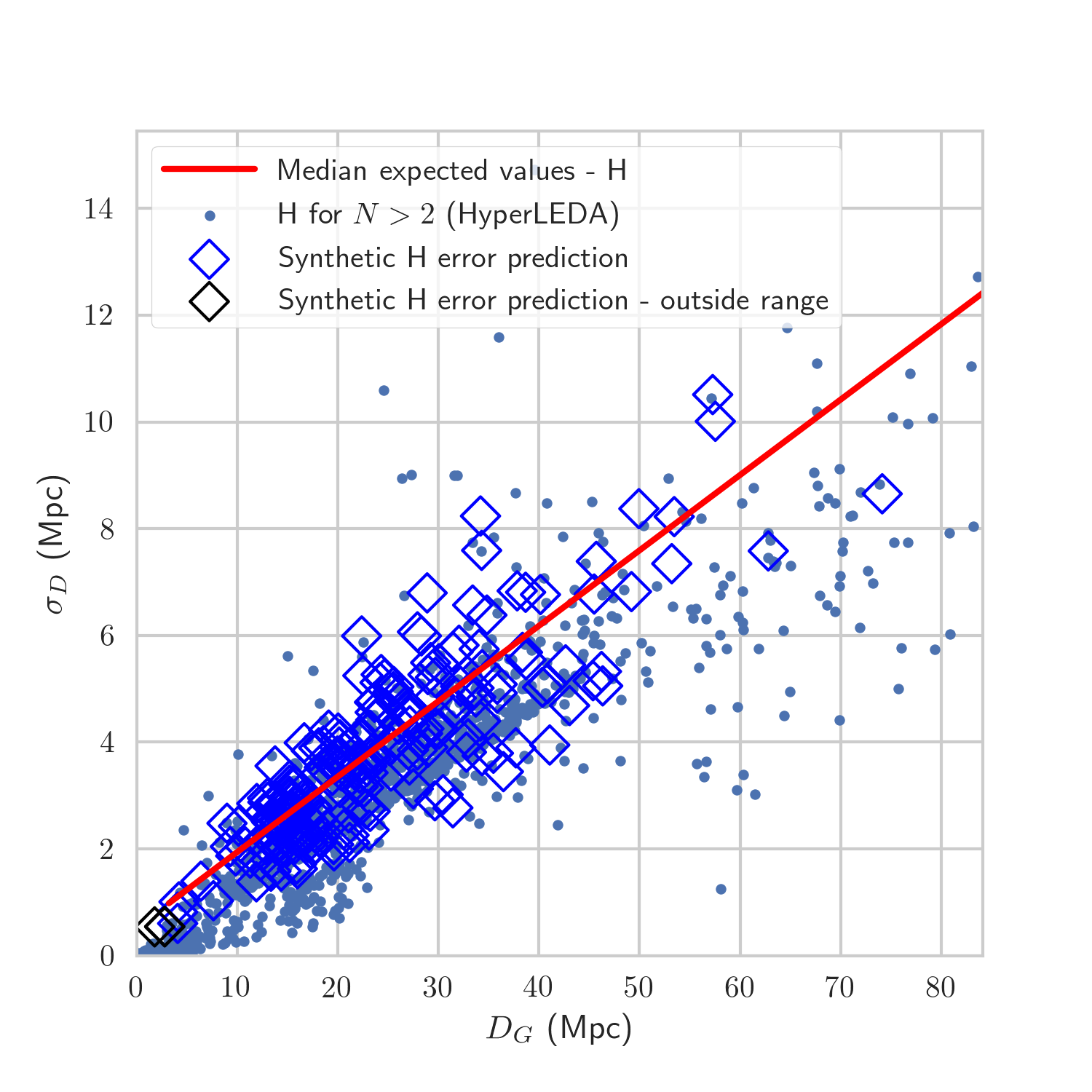}
	\includegraphics[scale=0.69]{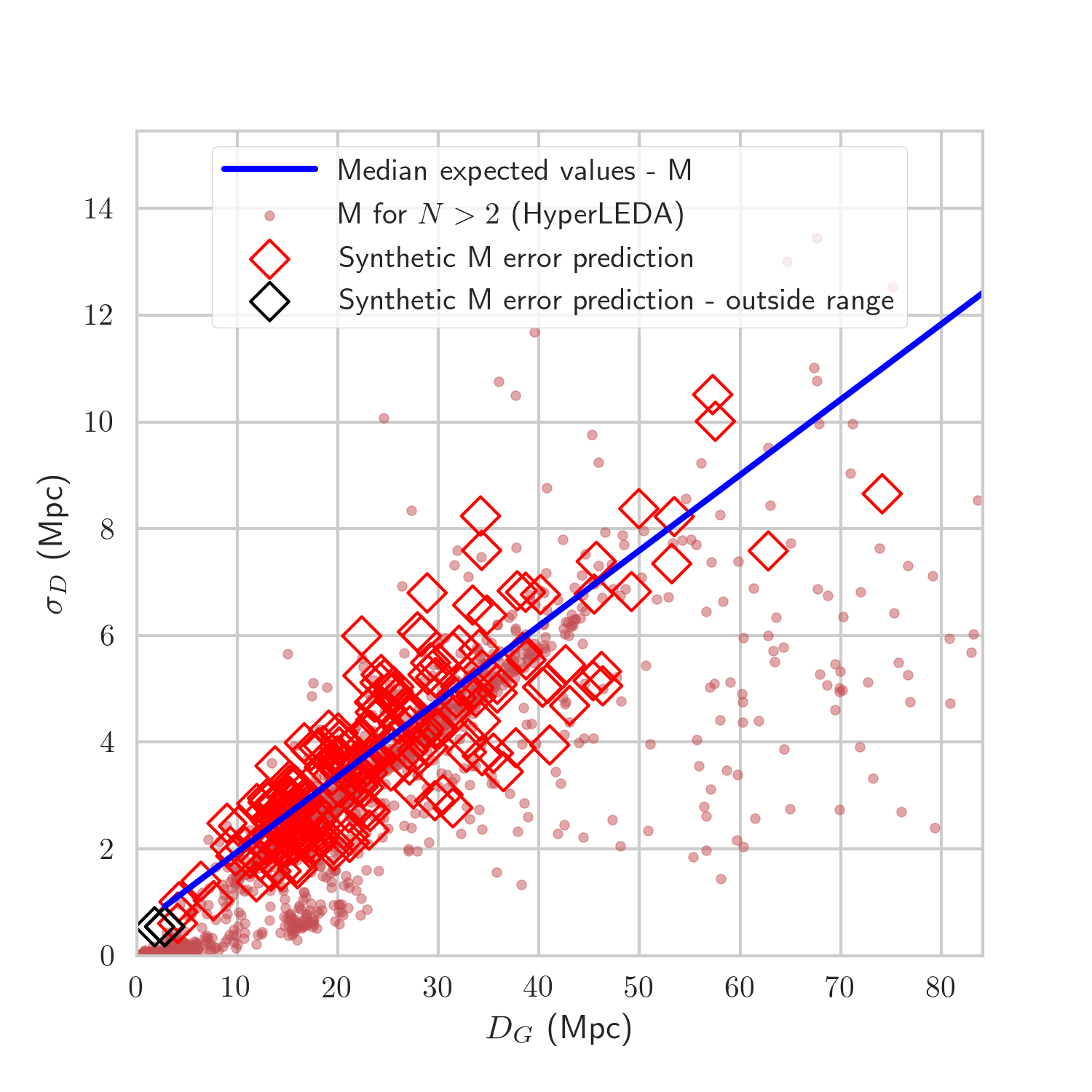}
    \caption{Synthetic H-method (left) and M-method (right) $\sigma_D$ and their median expected values vs. $D_G$ for the 71 galaxies in HyperLEDA for which no TF distance measurements report an error, generated using the corresponding Bayesian linear model. Predicted errors for galaxies outside of the working distance range of the model are plotted in black. H errors for galaxies with more than 2 distance measurements are also plotted for comparison.}
    \label{fig:predhl1}
\end{figure*}

The HyperLEDA catalog has distance measurements for 4224 galaxies, of which 1064 galaxies have reported measurements without errors. Of these galaxies with unreported distance errors, 203 report measurements obtained with the TF method. We create synthetic errors for these using our Bayesian predictive models for H and M TF errors. Fig.~\ref{fig:predhl1} (left) shows that predicted H errors are somewhat higher than those estimated for HyperLEDA, although acceptably within the range. Fig.~\ref{fig:predhl1} (right) shows that predicted M errors are even closer to the HyperLEDA M error estimates. This outstanding result is an independent validation of our linear Bayesian model for predicting TF distance errors, and its capacity to estimate systematic effects of the TF distance determination method.\\

This predictive model may work for other distance determination methods, but a cursory overview of methods which require error prediction due to missing errors (e.g. TRGB, CMD, Eclipsing Binary, Red Clump, PNLF, SZ effect, Brightest Stars, Horizontal Branch in NED-D) suggests that such attempts need to be evaluated in a case-by-case basis. For instance, in NED-D, Fundamental Plane (FP) measurements are by far the most numerous ($\sim130k$ galaxies), but only 28 of those have more than 3 FP distance measurements. We attempted to create a model similar to what we did for TF, but we were only able to find a working predictive model (i.e. yielding a good Bayesian $p$-value) for the 16 galaxies with more than 4 distance measurements. The comparatively low number of galaxies for which this model works makes us wary of predicting FP errors, therefore we do not report these results.

\section{Conclusions}

We propose methods for robustly estimating the uncertainty in extragalactic distances in multi-measurement, multi-method catalogs. First we propose to report 16th, 50th, and 84th percentiles of the bootstrap-sampled distance distribution for each galaxy. We also propose the use of the half-distance between the 84th and 16th percentiles (method H), and the median absolute deviation (method M) if the bootstrap-sampled distance distribution for each galaxy as straightforward measures of the uncertainty in extragalactic distances. Method H gives errors that faithfully measure the variance of the distance probability distribution, whereas traditional frequentist propagation-of-error methods fail to match this variance measure. On the other hand, method M should be used whenever a specific application requires to ignore outdated or possibly wrong outliers.\\

We produce error data tables using the robust (H, M) and frequentist (P, Q) methods for NED-D, HyperLEDA, and Cosmicflows-3, along with the 16th, 50th, and 84th percentiles of the bootstrap-sampled distance distribution for each galaxy in those catalogs. These tables can be found in the repository for this paper, located at \texttt{http://github.com/saint-germain/errorprediction}. A description and analysis for each catalog can be found in the appendices. We consider that these error tables should be a fundamental tool for future precision cosmology, catalog-wide studies, as it should be possible to quote errors according to the method that the reader considers most relevant for specific applications.\\

We create a Bayesian predictive model for TF distance errors in the NED-D catalog based on a Bayesian analysis of the systematic and random components in distance errors. We perform a posterior predictive check in the form of the computation of a Bayesian $p$-value based on simulated vs. observed discrepancies measured with the Freeman-Tukey statistic. Thus we create models which can reproduce the intrinsic variance of distance errors along with systematic zero-setting and scale factor components from the posterior predictive distribution of the models, using NED-D estimated H and M errors.\\

We use these models to predict H and M errors for 884 galaxies in NED-D which report TF distance measurements but do not report measurement errors. Our predictive models are independently validated against the HyperLEDA catalog by the agreement between our pre-computed H and M errors and our predictions for 203 galaxies in HyperLEDA with non-reported TF errors. Similar Bayesian predictive methods can be set up for other distance determination methods but with caveats, as model validation works better for methods for which there are many galaxies with a high number of distance measurements.\\ 

Finally, we want to advocate for the widespread use of discrepancy plots and their derived Bayesian $p$-values for  Bayesian model checking in astronomy, as inference is based on the model's ability to reproduce the original distribution of the data and not only on a relative comparison to other models.

\section*{Acknowledgements}

The authors would like to thank O. L. Ram\'irez-Su\'arez and J. E. Forero-Romero for their valuable input during the early stages of this work. This research has made use of the NASA/IPAC Extragalactic Database (NED), which is operated by the Jet Propulsion Laboratory, California Institute of Technology, under contract with the National Aeronautics and Space Administration.

%%%%%%%%%%%%%%%%%%%%%%%%%%%%%%%%%%%%%%%%%%%%%%%%%%



\bibliographystyle{mnras}
\bibliography{savedrecs} % if your bibtex file is called example.bib

\begin{thebibliography}{}
\makeatletter
\relax
\def\mn@urlcharsother{\let\do\@makeother \do\$\do\&\do\#\do\^\do\_\do\%\do\~}
\def\mn@doi{\begingroup\mn@urlcharsother \@ifnextchar [ {\mn@doi@}
  {\mn@doi@[]}}
\def\mn@doi@[#1]#2{\def\@tempa{#1}\ifx\@tempa\@empty \href
  {http://dx.doi.org/#2} {doi:#2}\else \href {http://dx.doi.org/#2} {#1}\fi
  \endgroup}
\def\mn@eprint#1#2{\mn@eprint@#1:#2::\@nil}
\def\mn@eprint@arXiv#1{\href {http://arxiv.org/abs/#1} {{\tt arXiv:#1}}}
\def\mn@eprint@dblp#1{\href {http://dblp.uni-trier.de/rec/bibtex/#1.xml}
  {dblp:#1}}
\def\mn@eprint@#1:#2:#3:#4\@nil{\def\@tempa {#1}\def\@tempb {#2}\def\@tempc
  {#3}\ifx \@tempc \@empty \let \@tempc \@tempb \let \@tempb \@tempa \fi \ifx
  \@tempb \@empty \def\@tempb {arXiv}\fi \@ifundefined
  {mn@eprint@\@tempb}{\@tempb:\@tempc}{\expandafter \expandafter \csname
  mn@eprint@\@tempb\endcsname \expandafter{\@tempc}}}

\bibitem[\protect\citeauthoryear{Barris \& Tonry}{Barris \&
  Tonry}{2004}]{ridsn}
Barris B.,  Tonry J.,  2004, \mn@doi [ASTROPHYSICAL JOURNAL] {10.1086/424871},
  613, L21

\bibitem[\protect\citeauthoryear{Bishop et~al.,}{Bishop
  et~al.}{2007}]{bishopft}
Bishop Y. M.~M.,  et~al., 2007, Discrete Multivariate Analysis: Theory and
  Practice.
Springer

\bibitem[\protect\citeauthoryear{Brooks, Catchpole  \& Morgan}{Brooks
  et~al.}{2000}]{brooks}
Brooks S.~P.,  Catchpole E.~A.,   Morgan B. J.~T.,  2000, Statistical Science,
  15, 357

\bibitem[\protect\citeauthoryear{Brugger}{Brugger}{1969}]{wstdev}
Brugger R.~M.,  1969, \mn@doi [The American Statistician]
  {10.1080/00031305.1969.10481865}, 23, 32

\bibitem[\protect\citeauthoryear{Chambert, Rotella  \& Higgs}{Chambert
  et~al.}{2014}]{ppcinf}
Chambert T.,  Rotella J.~J.,   Higgs M.~D.,  2014, \mn@doi [Ecology and
  Evolution] {10.1002/ece3.993}, 4, 1389

\bibitem[\protect\citeauthoryear{{Chaparro Molano}, {Restrepo Gait{\'a}n},
  {Cuervo Marulanda}  \& {Torres Arzayus}}{{Chaparro Molano}
  et~al.}{2018}]{chaparro18}
{Chaparro Molano} G.,  {Restrepo Gait{\'a}n} O.~A.,  {Cuervo Marulanda} J.~C.,
   {Torres Arzayus} S.~A.,  2018, in Revista Mexicana de Astronomia y
  Astrofisica Conference Series. pp 63--63

\bibitem[\protect\citeauthoryear{Courtois, Hoffman, Tully  \&
  Gottloeber}{Courtois et~al.}{2012}]{locunivcf}
Courtois H.~M.,  Hoffman Y.,  Tully R.~B.,   Gottloeber S.,  2012, \mn@doi
  [ASTROPHYSICAL JOURNAL] {10.1088/0004-637X/744/1/43}, 744

\bibitem[\protect\citeauthoryear{De~la Horra}{De~la Horra}{2008}]{chi2ms}
De~la Horra J.,  2008, \mn@doi [COMMUNICATIONS IN STATISTICS-THEORY AND
  METHODS] {10.1080/03610920701678976}, 37, 1412

\bibitem[\protect\citeauthoryear{Dhawan, Jha  \& Leibundgut}{Dhawan
  et~al.}{2018}]{hubsn2018}
Dhawan S.,  Jha S.~W.,   Leibundgut B.,  2018, \mn@doi [ASTRONOMY \&
  ASTROPHYSICS] {10.1051/0004-6361/201731501}, 609

\bibitem[\protect\citeauthoryear{Foreman-Mackey, Hogg, Lang  \&
  Goodman}{Foreman-Mackey et~al.}{2013}]{emcee}
Foreman-Mackey D.,  Hogg D.~W.,  Lang D.,   Goodman J.,  2013, \mn@doi
  [PUBLICATIONS OF THE ASTRONOMICAL SOCIETY OF THE PACIFIC] {10.1086/670067},
  125, 306

\bibitem[\protect\citeauthoryear{Freedman \& Madore}{Freedman \&
  Madore}{2010}]{hub2010}
Freedman W.~L.,  Madore B.~F.,  2010, in Blandford R.,  Faber S.,  van Dishoeck
  E.,   Kormendy J.,  eds, Annual Review of Astronomy and Astrophysics,
  Vol.~48, ANNUAL REVIEW OF ASTRONOMY AND ASTROPHYSICS, VOL 48.
pp 673--710, \mn@doi{10.1146/annurev-astro-082708-101829}

\bibitem[\protect\citeauthoryear{Freedman et~al.,}{Freedman
  et~al.}{2001}]{huborig}
Freedman W.~L.,  et~al., 2001, The Astrophysical Journal, 553, 47

\bibitem[\protect\citeauthoryear{Gelman}{Gelman}{2003}]{gelman2003}
Gelman A.,  2003, Internat. Statist. Rev., 71, 369

\bibitem[\protect\citeauthoryear{Gelman, li Meng  \& Stern}{Gelman
  et~al.}{1996}]{gelmanppd}
Gelman A.,  li Meng X.,   Stern H.,  1996, Statistica Sinica, 6, 733

\bibitem[\protect\citeauthoryear{Humphreys, Reid, Moran, Greenhill  \&
  Argon}{Humphreys et~al.}{2013}]{hubngc}
Humphreys E. M.~L.,  Reid M.~J.,  Moran J.~M.,  Greenhill L.~J.,   Argon A.~L.,
   2013, \mn@doi [ASTROPHYSICAL JOURNAL] {10.1088/0004-637X/775/1/13}, 775

\bibitem[\protect\citeauthoryear{Jarrett, Chester, Cutri, Schneider, Skrutskie
  \& Huchra}{Jarrett et~al.}{2000}]{2mass}
Jarrett T.~H.,  Chester T.,  Cutri R.,  Schneider S.,  Skrutskie M.,   Huchra
  J.~P.,  2000, The Astronomical Journal, 119, 2498

\bibitem[\protect\citeauthoryear{Javanmardi \& Kroupa}{Javanmardi \&
  Kroupa}{2017}]{morphanis}
Javanmardi B.,  Kroupa P.,  2017, \mn@doi [ASTRONOMY \& ASTROPHYSICS]
  {10.1051/0004-6361/201629408}, 597

\bibitem[\protect\citeauthoryear{{Jesus}, {Greg{\'o}rio}, {Andrade-Oliveira},
  {Valentim}  \& {Matos}}{{Jesus} et~al.}{2018}]{bayesh}
{Jesus} J.~F.,  {Greg{\'o}rio} T.~M.,  {Andrade-Oliveira} F.,  {Valentim} R.,
  {Matos} C.~A.~O.,  2018, \mn@doi [\mnras] {10.1093/mnras/sty813}, \href
  {http://adsabs.harvard.edu/abs/2018MNRAS.tmp..789J} {}

\bibitem[\protect\citeauthoryear{Kelly}{Kelly}{2007}]{gmastro}
Kelly B.~C.,  2007, \mn@doi [ASTROPHYSICAL JOURNAL] {10.1086/519947}, 665, 1489

\bibitem[\protect\citeauthoryear{Kourkchi \& Tully}{Kourkchi \&
  Tully}{2017}]{gg3500}
Kourkchi E.,  Tully R.~B.,  2017, \mn@doi [ASTROPHYSICAL JOURNAL]
  {10.3847/1538-4357/aa76db}, 843

\bibitem[\protect\citeauthoryear{Ma, Taylor  \& Scott}{Ma
  et~al.}{2013}]{nongauss}
Ma Y.-Z.,  Taylor J.~E.,   Scott D.,  2013, \mn@doi [MONTHLY NOTICES OF THE
  ROYAL ASTRONOMICAL SOCIETY] {10.1093/mnras/stt1726}, 436, 2029

\bibitem[\protect\citeauthoryear{Makarov, Prugniel, Terekhova, Courtois  \&
  Vauglin}{Makarov et~al.}{2014}]{hyperleda}
Makarov D.,  Prugniel P.,  Terekhova N.,  Courtois H.,   Vauglin I.,  2014,
  \mn@doi [ASTRONOMY \& ASTROPHYSICS] {10.1051/0004-6361/201423496}, 570

\bibitem[\protect\citeauthoryear{Mazzarella \& Team}{Mazzarella \&
  Team}{2007}]{ned07}
Mazzarella J.~M.,  Team N.,  2007, in Shaw R.,  Hill F.,   Bell D.,  eds,
  ASTRONOMICAL SOCIETY OF THE PACIFIC CONFERENCE SERIES Vol. 376, ASTRONOMICAL
  DATA ANALYSIS SOFTWARE AND SYSTEMS XVI. pp 153--162

\bibitem[\protect\citeauthoryear{McClure \& Dyer}{McClure \&
  Dyer}{2007}]{anishub}
McClure M.~L.,  Dyer C.~C.,  2007, \mn@doi [NEW ASTRONOMY]
  {10.1016/j.newast.2007.03.005}, 12, 533

\bibitem[\protect\citeauthoryear{Mould \& Sakai}{Mould \& Sakai}{2008}]{noceph}
Mould J.,  Sakai S.,  2008, \mn@doi [ASTROPHYSICAL JOURNAL LETTERS]
  {10.1086/592964}, 686, L75

\bibitem[\protect\citeauthoryear{Nasonova \& Karachentsev}{Nasonova \&
  Karachentsev}{2011}]{void}
Nasonova O.~G.,  Karachentsev I.~D.,  2011, \mn@doi [ASTROPHYSICS]
  {10.1007/s10511-011-9153-1}, 54, 1

\bibitem[\protect\citeauthoryear{Obreschkow \& Meyer}{Obreschkow \&
  Meyer}{2013}]{precisetf}
Obreschkow D.,  Meyer M.,  2013, \mn@doi [ASTROPHYSICAL JOURNAL]
  {10.1088/0004-637X/777/2/140}, 777

\bibitem[\protect\citeauthoryear{Riess et~al.,}{Riess et~al.}{2016}]{riess}
Riess A.~G.,  et~al., 2016, The Astrophysical Journal, 826, 56

\bibitem[\protect\citeauthoryear{Roman \& Trujillo}{Roman \&
  Trujillo}{2017}]{gallargescale}
Roman J.,  Trujillo I.,  2017, \mn@doi [MONTHLY NOTICES OF THE ROYAL
  ASTRONOMICAL SOCIETY] {10.1093/mnras/stx438}, 468, 703

\bibitem[\protect\citeauthoryear{Rubin et~al.,}{Rubin et~al.}{2015}]{unity}
Rubin D.,  et~al., 2015, \mn@doi [ASTROPHYSICAL JOURNAL]
  {10.1088/0004-637X/813/2/137}, 813

\bibitem[\protect\citeauthoryear{Said, Kraan-Korteweg, Staveley-Smith,
  Williams, Jarrett  \& Springob}{Said et~al.}{2016}]{said}
Said K.,  Kraan-Korteweg R.~C.,  Staveley-Smith L.,  Williams W.~L.,  Jarrett
  T.~H.,   Springob C.~M.,  2016, \mn@doi [MONTHLY NOTICES OF THE ROYAL
  ASTRONOMICAL SOCIETY] {10.1093/mnras/stw105}, 457, 2366

\bibitem[\protect\citeauthoryear{Sorce et~al.,}{Sorce et~al.}{2013}]{sorce}
Sorce J.~G.,  et~al., 2013, The Astrophysical Journal, 765, 94

\bibitem[\protect\citeauthoryear{Sorce, Courtois, Gottlober, Hoffman  \&
  Tully}{Sorce et~al.}{2014}]{localunipv}
Sorce J.~G.,  Courtois H.~M.,  Gottlober S.,  Hoffman Y.,   Tully R.~B.,  2014,
  \mn@doi [MONTHLY NOTICES OF THE ROYAL ASTRONOMICAL SOCIETY]
  {10.1093/mnras/stt2153}, 437, 3586

\bibitem[\protect\citeauthoryear{Speagle \& Eisenstein}{Speagle \&
  Eisenstein}{2017a}]{photred1}
Speagle J.~S.,  Eisenstein D.~J.,  2017a, \mn@doi [MONTHLY NOTICES OF THE ROYAL
  ASTRONOMICAL SOCIETY] {10.1093/mnras/stw1485}, 469, 1186

\bibitem[\protect\citeauthoryear{Speagle \& Eisenstein}{Speagle \&
  Eisenstein}{2017b}]{photred2}
Speagle J.~S.,  Eisenstein D.~J.,  2017b, \mn@doi [MONTHLY NOTICES OF THE ROYAL
  ASTRONOMICAL SOCIETY] {10.1093/mnras/stx510}, 469, 1205

\bibitem[\protect\citeauthoryear{Springob, Masters, Haynes, Giovanelli  \&
  Marinoni}{Springob et~al.}{2007}]{tf07dist}
Springob C.~M.,  Masters K.~L.,  Haynes M.~P.,  Giovanelli R.,   Marinoni C.,
  2007, The Astrophysical Journal Supplement Series, 172, 599

\bibitem[\protect\citeauthoryear{Springob et~al.,}{Springob et~al.}{2014}]{6df}
Springob C.~M.,  et~al., 2014, \mn@doi [MONTHLY NOTICES OF THE ROYAL
  ASTRONOMICAL SOCIETY] {10.1093/mnras/stu1743}, 445, 2677

\bibitem[\protect\citeauthoryear{Steer et~al.,}{Steer et~al.}{2017}]{ned}
Steer I.,  et~al., 2017, \mn@doi [ASTRONOMICAL JOURNAL]
  {10.3847/1538-3881/153/1/37}, 153

\bibitem[\protect\citeauthoryear{Torres \& Cuervo}{Torres \&
  Cuervo}{2018}]{tecciencia}
Torres S.,  Cuervo J.~C.,  2018, \mn@doi [Tecciencia]
  {10.18180/tecciencia.2018.24.2}, 24, 53

\bibitem[\protect\citeauthoryear{{Tully} \& {Fisher}}{{Tully} \&
  {Fisher}}{1977}]{tforig}
{Tully} R.~B.,  {Fisher} J.~R.,  1977, \aap, \href
  {http://adsabs.harvard.edu/abs/1977A%26A....54..661T} {54, 661}

\bibitem[\protect\citeauthoryear{Tully \& Pierce}{Tully \&
  Pierce}{2000}]{hubunc}
Tully R.~B.,  Pierce M.~J.,  2000, The Astrophysical Journal, 533, 744

\bibitem[\protect\citeauthoryear{Tully, Courtois  \& Sorce}{Tully
  et~al.}{2016}]{cosmicflows}
Tully R.~B.,  Courtois H.~M.,   Sorce J.~G.,  2016, \mn@doi [ASTRONOMICAL
  JOURNAL] {10.3847/0004-6256/152/2/50}, 152

\bibitem[\protect\citeauthoryear{Watkins \& Feldman}{Watkins \&
  Feldman}{2015}]{lognormal}
Watkins R.,  Feldman H.~A.,  2015, \mn@doi [Monthly Notices of the Royal
  Astronomical Society] {10.1093/mnras/stv651}, 450, 1868

\bibitem[\protect\citeauthoryear{White, Daw  \& Dhillon}{White
  et~al.}{2011}]{gwgallist}
White D.~J.,  Daw E.~J.,   Dhillon V.~S.,  2011, \mn@doi [CLASSICAL AND QUANTUM
  GRAVITY] {10.1088/0264-9381/28/8/085016}, 28

\bibitem[\protect\citeauthoryear{Zhang \& Shields}{Zhang \&
  Shields}{2018}]{propprob2018}
Zhang J.,  Shields M.~D.,  2018, \mn@doi [MECHANICAL SYSTEMS AND SIGNAL
  PROCESSING] {10.1016/j.ymssp.2017.04.042}, 98, 465

\bibitem[\protect\citeauthoryear{de~la Horra \& Teresa Rodriguez-Bernal}{de~la
  Horra \& Teresa Rodriguez-Bernal}{2012}]{otherdisc}
de~la Horra J.,  Teresa Rodriguez-Bernal M.,  2012, SORT-STATISTICS AND
  OPERATIONS RESEARCH TRANSACTIONS, 36, 69

\makeatother
\end{thebibliography}

%\begin{thebibliography}{99}
%\end{thebibliography}

%%%%%%%%%%%%%%%%%%%%%%%%%%%%%%%%%%%%%%%%%%%%%%%%%%

%%%%%%%%%%%%%%%%% APPENDICES %%%%%%%%%%%%%%%%%%%%%

\appendix

\section{Pre-computed distance error data tables}
We estimated errors for the HyperLEDA, Cosmicflows-3, and NED-D redshift-independent extragalactic distance databases using the methods described in Section~\ref{sec:comp} across all distance determination methods, only considering measurements with more than two reported errors. Our pre-computed distance error tables can be found in the repository for this paper at \texttt{http://github.com/saint-germain/errorprediction}. The fields included for each catalog are:
\begin{itemize}
\item \texttt{meas} - Number of distance measurements.
\item \texttt{D (Mpc)} - This is the median of the posterior distribution of the corresponding extragalactic distance.
\item \texttt{Dmin (Mpc)} - This is the 16th percentile of the posterior distribution of the corresponding extragalactic distance.
\item \texttt{Dmin (Mpc)} - This is the 84th percentile of the posterior distribution of the corresponding extragalactic distance.
\item \texttt{H (Mpc)} - Error estimated using the H method (\texttt{Dmax-Dmin})/2.
\item \texttt{M (Mpc)} - Error estimated using the M method. 
\item \texttt{P (Mpc)} - Error estimated using the P method. 
\item \texttt{Q (Mpc)} - Error estimated using the Q method. 
\end{itemize}

\subsection{Estimation of errors for HyperLEDA}
\begin{figure*}
	\includegraphics[scale=0.69]{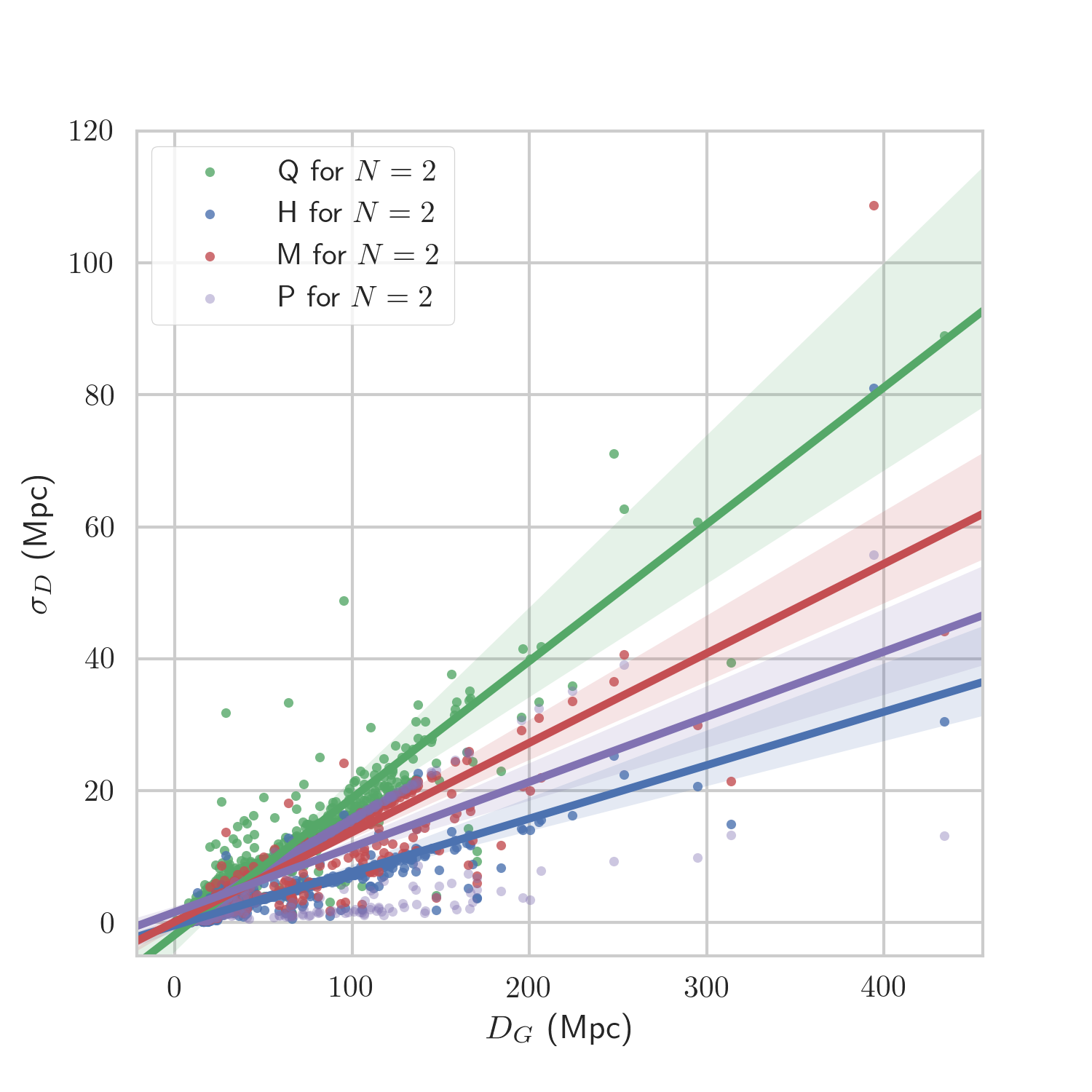}
	\includegraphics[scale=0.69]{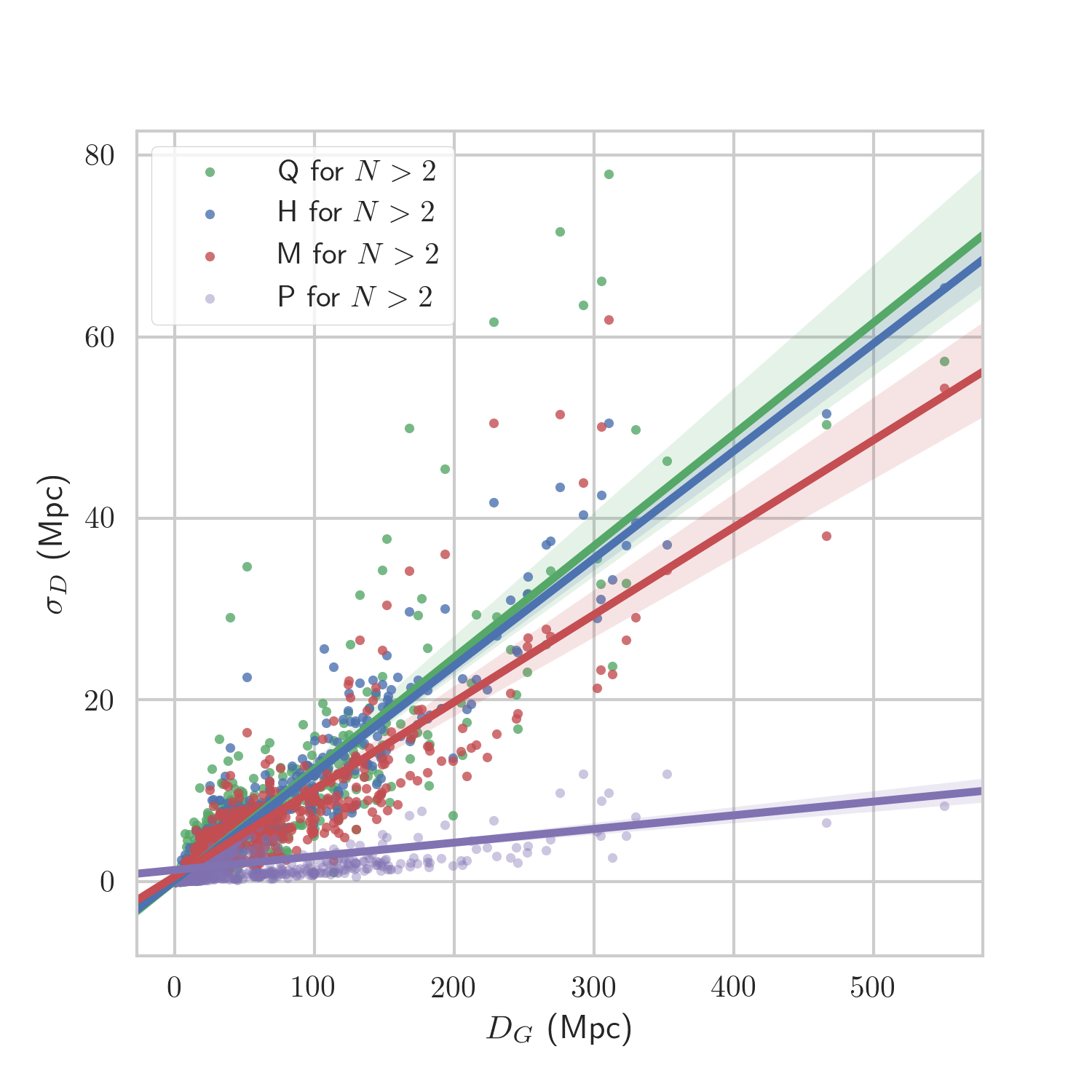}
    \caption{Estimated extragalactic distance errors vs. median extragalactic distance for galaxies with $N=2$ (left) and $N>2$ (right) redshift-independent distance measurements in HyperLEDA according to the H, M, Q, P error models, showing a linear regression and confidence intervals computed using the \texttt{seaborn.regplot} Python function.}
    \label{fig:HLlow}
\end{figure*}

As expected from our analysis of TF errors in NED-D, errors calculated with methods P, Q, and M overpredict the error with respect to the H method for galaxies with a low number of distance measurements ($N=2$), as shown in Fig.~\ref{fig:HLlow} (left). Fig.~\ref{fig:HLlow} (right) shows that for galaxies with a higher number of distance measurements, the P method significantly underpredicts the error with respect to the other methods. Even though the H and Q methods show a similar trend, the variance of Q errors around this trend is higher than for H methods. Errors obtained with method M are lower, due to the method's intrinsic robustness. These estimations are reported in the file called \texttt{hl\_bootstrap\_results.csv} in the repository. Special fields for this catalog are:
\begin{itemize}
\item \texttt{objname} - Object name according to the HyperLEDA database.
\item \texttt{j2000} - J2000 coordinates.
\end{itemize}

\subsection{Estimation of errors for Cosmicflows-3}
\begin{figure*}
	\includegraphics[scale=0.69]{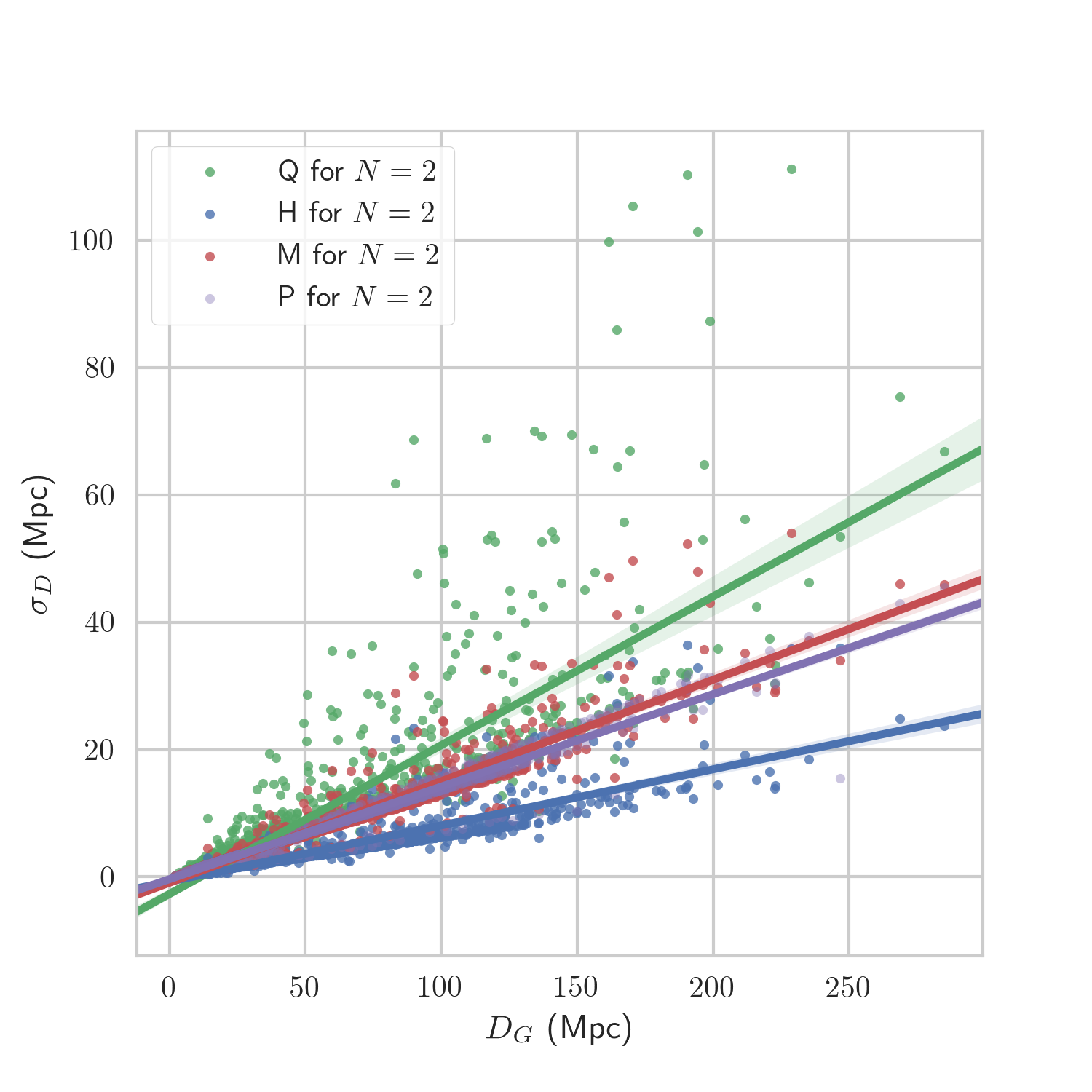}
	\includegraphics[scale=0.69]{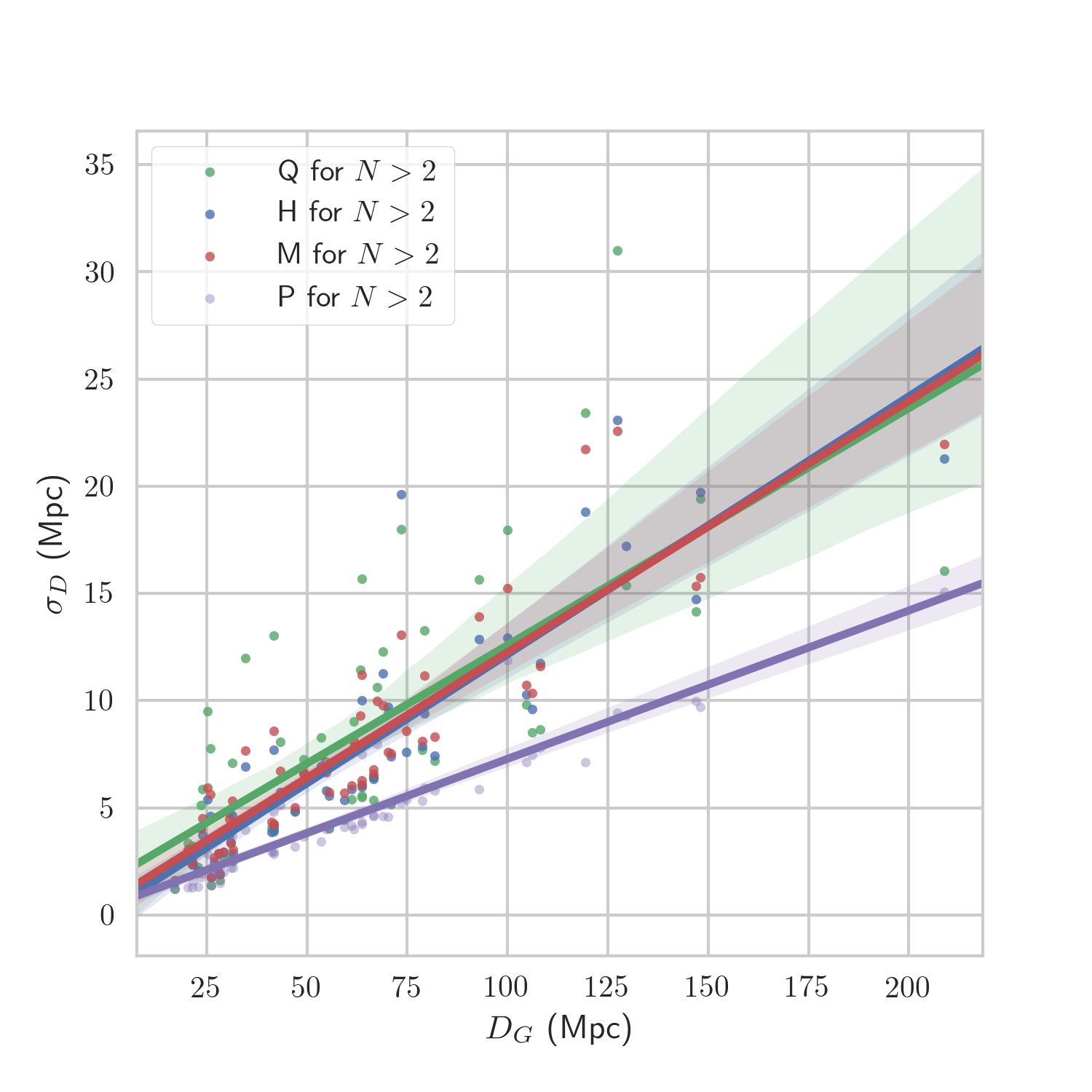}
    \caption{Estimated extragalactic distance errors vs. median extragalactic distance for galaxies with $N=2$ (left) and $N>2$ (right) redshift-independent distance measurements in Cosmicflows-3 according to the H, M, Q, P error models, showing a linear regression and confidence intervals computed using the \texttt{seaborn.regplot} Python function.}
    \label{fig:CF3low}
\end{figure*}

The Extragalactic Distance Database (EDD) of Cosmicflows-3, which has the most up-to-date calibrated distance measurements using the TF, FP, SNIa methods for more than 17000 galaxies. We estimated errros for the approximately 10\% of them which have more than one reported distance, using the methods described in Section~\ref{sec:comp}. Fig.~\ref{fig:CF3low} shows that the P method, which is the suggested method in \citet{cosmicflows}, overpredicts errors with respect to the H method for galaxies with 2 distance measurements, as was the case for the errors of HyperLEDA. For galaxies with more than 2 distance measurements, Fig.~\ref{fig:CF3low} shows that the P method underpredicts the errors with respect to the H method. Even though the M, H, and Q methods show a similar trend with distance, the Q method has a significantly larger scatter around this trend. We compiled the estimated errors in a companion table to the Cosmicflows-3 EDD database and in a similar format, in the file called \texttt{cf3\_bootstrap\_results.csv} in the repository for this work. Special fields for this catalog are:
\begin{itemize}
\item \texttt{pgc} - Principal Galaxies Catalog ID number.
\item \texttt{Name} - Object name according to the Cosmicflows-3 database, where available.
\end{itemize}

\subsection{Estimation of errors for NED-D}
The 2018 version of the NASA/IPAC extragalactic distance catalog NED-D has $\sim300000$ redshift-independent distance measurements with reported errors for $\sim180000$ galaxies. We estimated the errors for the $\sim16000$ galaxies with more than one distance measurement. The database of errors for NED-D is in the file called \texttt{ned\_bootstrap\_results.csv}. The only special field for this catalog is:
\begin{itemize}
\item \texttt{Galaxy ID} - Object name according to the NED-D database.
\end{itemize}

%%%%%%%%%%%%%%%%%%%%%%%%%%%%%%%%%%%%%%%%%%%%%%%%%%

%\citep[e.g.][]{photred1}.
%Figures are referred to as e.g. Fig.~\ref{fig:example_figure}, and tables as
%e.g. Table~\ref{tab:example_table}.
%\begin{table}
%	\centering
%	\caption{This is an example table. Captions appear above each table.
%	Remember to define the quantities, symbols and units used.}
%	\label{tab:example_table}
%	\begin{tabular}{lccr} % four columns, alignment for each
%		\hline
%		A & B & C & D\\
%		\hline
%		1 & 2 & 3 & 4\\
%		2 & 4 & 6 & 8\\
%		3 & 5 & 7 & 9\\
%		\hline
%	\end{tabular}
%\end{table}

% Don't change these lines
\bsp	% typesetting comment
\label{lastpage}
\end{document}